\documentstyle[11pt,aaspp4]{article}
\begin{document}

\slugcomment{To be published in the Astronomical Journal, Dec 2001}

\title{A Survey of $z>5.8$ Quasars in the Sloan Digital Sky Survey I:
Discovery of Three New Quasars and the Spatial Density of Luminous
Quasars at $z\sim 6$\altaffilmark{1,2}}

\author{Xiaohui Fan\altaffilmark{\ref{IAS}},
Vijay K. Narayanan\altaffilmark{\ref{Princeton}},
Robert H. Lupton\altaffilmark{\ref{Princeton}},
Michael A. Strauss\altaffilmark{\ref{Princeton}},
Gillian R. Knapp\altaffilmark{\ref{Princeton}},
Robert H. Becker\altaffilmark{\ref{UCDavis},\ref{IGPP}},
Richard L. White\altaffilmark{\ref{STScI}},
Laura Pentericci\altaffilmark{\ref{Heidelberg}},
S. K. Leggett\altaffilmark{\ref{JAC}},
Zolt\'an Haiman\altaffilmark{\ref{Princeton},\ref{Hubble}},
James E. Gunn\altaffilmark{\ref{Princeton}},
\v{Z}eljko Ivezi\'{c}\altaffilmark{\ref{Princeton}},
Donald P. Schneider\altaffilmark{\ref{PSU}},
Scott F. Anderson\altaffilmark{\ref{Washington}},
J. Brinkmann\altaffilmark{\ref{APO}},
Neta A. Bahcall\altaffilmark{\ref{Princeton}},
Andrew J. Connolly\altaffilmark{\ref{Pitt}},
Istvan Csabai\altaffilmark{\ref{JHU},\ref{Eotvos}},
Mamoru Doi\altaffilmark{\ref{UTokyo}},
Masataka Fukugita\altaffilmark{\ref{CosmicRay}},
Tom Geballe\altaffilmark{\ref{Gemini}},
Eva K. Grebel\altaffilmark{\ref{Heidelberg}},
Daniel Harbeck\altaffilmark{\ref{Heidelberg}},
Gregory Hennessy\altaffilmark{\ref{USNO}},
Don Q. Lamb\altaffilmark{\ref{Chicago}},
%Pete Newman\altaffilmark{\ref{APO}},
Gajus Miknaitis\altaffilmark{\ref{Washington}},
Jeffrey A. Munn\altaffilmark{\ref{Flagstaff}},
Robert Nichol\altaffilmark{\ref{CMU}},
Sadanori Okamura\altaffilmark{\ref{UTokyo}},
Jeffrey R. Pier\altaffilmark{\ref{Flagstaff}},
Francisco Prada\altaffilmark{\ref{CalarAlto}},
Gordon T. Richards\altaffilmark{\ref{PSU}},
Alex Szalay\altaffilmark{\ref{JHU}},
Donald G. York\altaffilmark{\ref{Chicago}}
}

\altaffiltext{1}{Based on observations obtained with the
Sloan Digital Sky Survey,
and with the Apache Point Observatory
3.5-meter telescope,
which is owned and operated by the Astrophysical Research Consortium;
on observations obtained by staff of the Gemini Observatory, 
which is operated by the Association of
Universities for Research in Astronomy, Inc., 
under a cooperative agreement with the NSF on behalf of the
Gemini partnership: The National Science Foundation (United States), the Particle Physics and Astronomy
Research Council (United Kingdom), the National Research Council (Canada), CONICYT (Chile), the
Australian Research Council (Australia), CNPq (Brazil), and CONICET (Argentina); 
on observations obtained at the W.M. Keck Observatory, which is operated as a scientific partnership
among the California Institute of Technology, the University of California and the National Aeronautics and
Space Administration, made possible by the generous financial support of the W.M. Keck
Foundation;
on observations obtained at the German-Spanish Astronomical Centre, Calar Alto 
Observatory,
operated by the Max Planck Institute for Astronomy, Heidelberg jointly with the Spanish National Commission for
Astronomy; and on observations obtained at UKIRT, which 
is operated by the Joint Astronomy Centre on behalf of the UK
Particle Physics and Astronomy Research Council.}
\altaffiltext{2}{
This paper is dedicated to the memory of Arthur F. Davidsen (1944-2001), a
pioneer of the study of the intergalactic medium, and a leader in the
development of the Sloan Digital Sky Survey.
}
\newcounter{address}
\setcounter{address}{3}
\altaffiltext{\theaddress}{Institute for Advanced Study, Olden Lane,
Princeton, NJ 08540
\label{IAS}}
\addtocounter{address}{1}
\altaffiltext{\theaddress}{Princeton University Observatory, Princeton,
NJ 08544
\label{Princeton}}
\addtocounter{address}{1}
\altaffiltext{\theaddress}{Physics Department, University of California, Davis,
CA 95616
\label{UCDavis}}
\addtocounter{address}{1}
\altaffiltext{\theaddress}{IGPP/Lawrence Livermore National Laboratory, Livermore,
CA 95616
\label{IGPP}}
\addtocounter{address}{1}
\altaffiltext{\theaddress}{Space Telescope Science Institute, Baltimore, MD 21218
\label{STScI}}
\addtocounter{address}{1}
\altaffiltext{\theaddress}{Max-Planck-Institut f\"{u}r Astronomie,
K\"{o}nigstuhl 17, D-69171 Heidelberg, Germany
\label{Heidelberg}}
\addtocounter{address}{1}
\altaffiltext{\theaddress}{United Kingdom Infrared Telescope, Joint Astronomy Center,
660 North A'ohoku Place, Hilo, Hawaii 96720
\label{JAC}}
\addtocounter{address}{1}
\altaffiltext{\theaddress}{Department of Astronomy and Astrophysics,
The Pennsylvania State University,
University Park, PA 16802
\label{PSU}}
\addtocounter{address}{1}
\altaffiltext{\theaddress}{Hubble Fellow\label{Hubble}}
\addtocounter{address}{1}
\altaffiltext{\theaddress}{University of Washington, Department of Astronomy,
Box 351580, Seattle, WA 98195
\label{Washington}}
\addtocounter{address}{1}
\altaffiltext{\theaddress}{Apache Point Observatory, P. O. Box 59,
Sunspot, NM 88349-0059
\label{APO}}
\newpage
\addtocounter{address}{1}
\altaffiltext{\theaddress}{
Department of Physics and Astronomy, University of Pittsburgh, Pittsburgh,
PA 15260
\label{Pitt}}
\addtocounter{address}{1}
\altaffiltext{\theaddress}{
Department of Physics and Astronomy, The Johns Hopkins University,
   3701 San Martin Drive, Baltimore, MD 21218, USA
\label{JHU}}
\addtocounter{address}{1}
\altaffiltext{\theaddress}{Department of Physics of Complex Systems,
E\"{o}tv\"{o}s University, Budapset, H-1117, Hungary
\label{Eotvos}}
\addtocounter{address}{1}
\altaffiltext{\theaddress}{Department of Astronomy and Research Center
  for the Early Universe, School of Science, University of Tokyo, Hongo,
  Bunkyo, Tokyo, 113-0033, Japan
\label{UTokyo}}
\addtocounter{address}{1}
\altaffiltext{\theaddress}{Institute for Cosmic Ray Research, University of
Tokyo, Midori, Tanashi, Tokyo 188-8502, Japan
\label{CosmicRay}}
\addtocounter{address}{1}
\altaffiltext{\theaddress}{Gemini Observatory, 670 North A'ohoku Place, Hilo,
HI 96720
\label{Gemini}}
\addtocounter{address}{1}
\altaffiltext{\theaddress}{U.S. Naval Observatory,
3450 Massachusetts Ave., NW,
Washington, DC  20392-5420
\label{USNO}}
\addtocounter{address}{1}
\altaffiltext{\theaddress}{University of Chicago, Astronomy \& Astrophysics
Center, 5640 S. Ellis Ave., Chicago, IL 60637
\label{Chicago}}
\addtocounter{address}{1}
\altaffiltext{\theaddress}{U.S. Naval Observatory, Flagstaff Station,
P.O. Box 1149,
Flagstaff, AZ  86002-1149
\label{Flagstaff}}
\addtocounter{address}{1}
\altaffiltext{\theaddress}{
Department of Physics, Carnegie Mellon University, Pittsburgh, PA 15213
\label{CMU}}
\addtocounter{address}{1}
\altaffiltext{\theaddress}{Centro Astron\'omico Hispano Alem\'an,
C/Jes\'us Durb\'an Rem\'on 2--2, E-04004 Almer\'\i{}a, Spain
\label{CalarAlto}}

\begin{abstract}

 We present the results from a survey of $i$-dropout objects 
selected from $\sim 1550$ deg$^2$ of multicolor imaging data
from the Sloan Digital Sky Survey (SDSS), to search for 
luminous quasars at $z\gtrsim 5.8$. 
Objects with $i^*-z^* > 2.2$ and $z^* < 20.2$ are selected, and follow-up $J$ 
band photometry is used to separate L and T type cool dwarfs from 
high-redshift quasars.
We describe the discovery of three new quasars, 
SDSSp J083643.85+005453.3 ($z=5.82$), SDSSp J130608.26+035626.3
($z=5.99$) and SDSSp J103027.10+052455.0 ($z=6.28$).
The quasar SDSSp J083643.85+005453.3 is a radio source with flux of
1.1$\,$mJy at 20$\,$cm.
The spectra of all three quasars show strong and broad Ly$\alpha$+NV emission lines, and 
very strong Ly$\alpha$ forest absorption, with a
mean continuum decrement $D_A > 0.90 $.
The ARC 3.5m spectrum of SDSSp J103027.10+052455.0 shows that 
over a range of $\sim 300$\AA\ immediately blueward of the Ly$\alpha$ 
emission,  the average transmitted flux is
only $0.003 \pm 0.020$ times that of the continuum level, 
consistent with zero flux over a $\sim 300$\AA\ range of the
Ly$\alpha$ forest region, and
suggesting a tentative detection of the complete Gunn-Peterson trough.  
%caused by neutral hydrogen in the intergalactic medium.
%The line ratio NV/CIV in the spectrum of this quasar
%suggests super-solar metallicity in the quasar environment.
The existence of strong metal lines in the quasar spectra 
suggests early metal enrichment in the quasar environment.

The three new objects, together with the previously published $z=5.8$ quasar
SDSSp J104433.04-012502.2, form a complete color-selected flux-limited 
sample at $z \gtrsim 5.8$.
We estimate the selection function of this sample, 
taking into account the estimated variations in the quasar spectral 
energy distribution as well as observational photometric errors.
We find that at $z=6$, 
the comoving density of luminous quasars at $M_{1450} < -26.8$ 
($H_0 = 50$ km s$^{-1}$ Mpc$^{-1}$, $\Omega = 1$) 
is $1.1 \times 10^{-9}$ Mpc$^{-3}$. 
This is a factor of $\sim 2$ lower than that at $z \sim 5$, 
and is consistent with an extrapolation of the observed quasar evolution 
at $z<5$.
%Assuming that these four quasars are not significantly 
%magnified by gravitational lensing, we show that the bright end slope 
%of the quasar luminosity function is flatter than $\Psi(L) \propto L^{-4.6}$,
%and discuss the implications for the contribution of quasars to the ionizing 
%background at $z \sim 6$. 
Using the current sample, we discuss the constraint on the
shape of the quasar luminosity function and the implications
for the contribution of quasars to the ionizing background
at $z\sim6$. 
The luminous quasars discussed in the paper have central black hole
masses of several times $10^9 \rm M_{\odot}$ by the Eddington argument, 
with likely dark halo masses of order $10^{13} \rm M_{\odot}$.
Their observed space density provides a sensitive test of models of quasar 
and galaxy formation at high redshift.

\end{abstract}


\section{Introduction}

High-redshift quasars provide direct probes of the epoch
when the first generation of galaxies and quasars formed.
The absorption spectra of these quasars reveal the state of the intergalactic 
medium (IGM) close to the reionization epoch 
(\cite{HL99}, \cite{MHR00}, \cite{MR00}, \cite{Cen00}).
The lack of a Gunn-Peterson trough (Shklovsky 1964, Scheuer 1965, 
Gunn \& Peterson 1965) in the spectrum of the luminous quasar 
SDSSp J104433.04--012502.2\footnote{The naming convention for the SDSS 
sources is SDSSp JHHMMSS.SS$\pm$DDMMSS.S, where ``p'' stands for the 
preliminary SDSS astrometry, and the positions are expressed in 
J2000.0 coordinates. The astrometry is accurate to better than $0.2''$
in each coordinate.} at $z=5.8$ (SDSS 1044--0125 for brevity, 
\cite{Fan00c}, for an updated redshift of this object, see
\cite{Goodrich01} and \cite{Djorgovski01}) indicates that the universe was already highly ionized
at that redshift.
Assuming that SDSS 1044-0125 is radiating at the Eddington luminosity, 
this object contains a central black hole of several billion solar masses.
The assembly of such massive objects in a timescale shorter than
1 Gyr yields constraints on models of the formation of massive black holes 
(e.g., \cite{HL01}).
The abundance and evolution of such quasars can provide sensitive
tests for models of quasar and galaxy evolution.

The Sloan Digital Sky Survey (SDSS;
\cite{York00,EDR}) uses
a dedicated 2.5m telescope and a large format CCD camera (\cite{Gunnetal})
at the Apache Point Observatory in New Mexico
to obtain images in five broad bands ($u$, $g$, $r$, $i$ and $z$
\footnote{Following \cite{EDR}, we refer to the SDSS passbands as
$u$, $g$, $r$, $i$ and $z$. As the SDSS photometric calibration
system is still being finalized, the SDSS photometry presented here
is referred to as $u^*$, $g^*$, $r^*$, $i^*$ and $z^*$.}
centered at 3551, 4686, 6166, 7480 and 8932 \AA, respectively; \cite{F96})
over 10,000 deg$^2$ of high Galactic latitude sky.
Over 200 quasars at $z>3.5$ have been discovered to date from the
SDSS multicolor imaging data 
(Fan et al.~1999a, 2000a,c, 2001a, \cite{Zheng00}, Schneider et
al. 2000, 2001, \cite{Anderson01}). 
The inclusion of the reddest band, $z$, in principle enables the
discovery of quasars up to $ z \sim 6.5$ from the SDSS data (Fan et
al.~2000c).  

Encouraged by the discovery of SDSS 1044--0125, we have carried out
a systematic survey of $i$-dropout quasars ($z\gtrsim 5.8$, whose
Lyman $\alpha$ forest is entirely in the $i$ band) based on
the SDSS imaging data. The goals of this survey are:
(1) to define a color-selected flux-limited complete sample of quasars at $z\sim 6$
in order to study the evolution of the quasar spatial density at high redshift;
(2) to provide multiple lines of sight for studying the nature of 
the high-redshift IGM;
(3) to study the intrinsic properties of quasars and the metallicity of
the quasar environment at high redshift.

In this paper, we present the first results of this survey, covering an area
of 1550 deg$^2$.
In \S 2, we present our color selection procedures for $z>5.8$ quasars.
Special software is developed to control the number of false
single-band detections in the imaging data. Follow-up optical and near-infrared photometry is used to 
separate high-redshift quasar candidates from the more numerous cool dwarfs 
which have similar optical colors. 
In \S 3, we describe the discovery of three new quasars at $z=5.82$, 
5.99 and 6.28. %, respectively, from the survey. 
We discuss their spectral properties and the implications
for the metallicity of the quasar environment and the status of the IGM in \S 4.
%Only the discovery spectra obtained with the 3.5m ARC telescope is presented here.
%High signal-to-noise ratio (S/N) spectra, obtained with Keck and VLT telescopes,
%and the accurate measurement
%of the quasar redshift and the emission and absorption properties of the new quasars, are presented in Paper II of this series.
These three quasars, combined with SDSS 1044--0125, form a complete
color-selected sample at $z^* < 20.2$. 
We calculate the survey selection function and derive the spatial density of 
$z\sim6$ quasars in \S 5.
In \S 6, we discuss the cosmological implications of the observations.
Follow-up spectroscopic observations carried out with the Keck
telescope will be presented by Becker et al.~(2001; Paper II). 

We present our results using two different cosmologies throughout the paper:
a $\Lambda$-dominated universe with $H_{0} = 65$ km s$^{-1}$ Mpc$^{-1}$,
$\Lambda = 0.65$ and $\Omega = 0.35$ (\cite{OS95}, \cite{KT95}), which
is referred to  as the $\Lambda$-model; and
a Einstein-de Sitter universe with 
$\Omega=1$ and $H_{0}$ = 50 km s$^{-1}$ Mpc$^{-1}$, which we refer to as
the $\Omega=1$ model in this paper.
The age of the universe is 13.9 Gyr and 13.0 Gyr in the 
$\Lambda$-model and the $\Omega=1$ model, respectively.

\section{Candidate Selection}

 \subsection{Colors of $z>5.8$ quasars}

The observed colors of quasars evolve strongly with redshift, as
the intrinsic emission and the intervening absorption spectral features 
move through the SDSS filter system (\cite{F99}, \cite{Richards01}).
The strong Ly$\alpha$ forest absorption blueward of the Ly$\alpha$ emission 
line enters the SDSS $g$, $r$ and $i$ bands at redshifts of about 
3.6, 4.6 and 5.5, respectively. At $z\gtrsim 5.7$, 
the Ly$\alpha$ emission line begins to move out of the SDSS
$i$ filter.
With an average $i^* - z^* \gtrsim 2$,
these high-redshift quasars 
have undetectable flux in the bluer $u$, $g$ and $r$ bands,
and little flux (sometimes also undetectable) in $i$ band.
They become $i$-dropout objects with only one measurable color
in SDSS photometry.

Figure 1 presents the $i^* - z^*$ vs. $z^*$ color-magnitude diagram
for 50,000 stellar sources selected at random from the SDSS imaging data.
It also shows the locations of all the $i$-dropout objects
(including L/T dwarfs and high-redshift quasars, see below)
selected in the 1550 deg$^2$ survey area.
The median track of simulated quasars (see \S5.1)
 with absolute magnitude $M_{1450} = -27$, representative
of the quasars discussed in this paper, is also plotted.
A simple cut at $i^* - z^* > 2.2$ selects quasar candidates at $z>5.8$.
At $z > 6.0$, the Ly$\alpha$ forest absorption enters the $z$ filter,
and the $z^*$ flux decreases rapidly with redshift with increasing amount
of absorption, while $i^* - z^*$ remains approximately constant, as
the increasing absorption in the $i$ and $z$ bands are roughly equal.
At $z \sim 6.6$, the Ly$\alpha$ emission line has redshifted to 9240\AA,
well past the effective wavelength of the $z$ filter.
Most of the $z$ flux is absorbed by the Ly$\alpha$ forest, and the object
disappears from optical images altogether.
This is the upper limit to the redshift of objects detectable with SDSS 
imaging data.

Quasars at $z>5.8$ are extremely rare on the sky.
We found four $i$-dropout quasars in 1550 deg$^2$.
For comparison, over the same area, there are about 15 million 
objects detected in the $z$ band above 6-$\sigma$, and about 6.5 million 
cosmic ray hits in the $z$ band. 
% at the 6-$\sigma$ level.
The quasars are faint, with low signal-to-noise ratio (S/N) photometry 
from the SDSS imaging data. 
Therefore, the key to a successful selection process is an efficient
elimination of potential contaminants.
We face three technical challenges:

(1) Elimination of false $z$-band only detections, composed primarily of
cosmic rays, but also including satellite trails, electronic ghost images 
and bleed trails from bright stars. 
While the recognition of the majority of cosmic rays  is relatively 
straightforward, even a tiny fraction  of mis-classified cosmic rays 
in the $z$ band, which would then appear as $i$-dropout candidates, 
would dominate over the much rarer true quasars. 
%be an overwhelming number of contaminants in our quasar survey 
%when an area of 1500 deg$^2$ is searched.
Note that SDSS scans the sky only once in the course of the survey except
in a small fraction of overlapping area; thus repeat observations
cannot be used to recognize cosmic rays.
Moreover, SDSS uses thick CCDs for the $z$ band.
A cosmic ray hit on these thick devices typically occupies more than a single
pixel, making the separation of real detections from cosmic ray 
hits more difficult, especially when the seeing is good.

(2) Reliability of $z^*$ photometry.
As is evident from Figure 1, the $i$-dropout objects are distributed on the
extreme tail of the $i^* - z^*$ distribution. Under typical seeing, $\sigma(z^*) \sim 0.1$
for objects with $z^* \sim 20$. 
Objects in the tail of the error distribution could scatter into
the selected region of color space, and pose a serious contamination
when a large number of objects is searched.
%In \S 2.2, we discuss the selection of ``real'' $i$-dropout candidates.

(3) Separating quasars and cool dwarfs.
As shown in \S 3, the surface density of cool dwarfs, with spectral types
ranging from mid-L to T, which have $i^* - z^* > 2.2$ (Geballe et al.~2001, Leggett et al.~2001), 
is $\sim 15$ times higher than that of $z>5.8$ quasars. 
We use $J$-band photometry to separate these two classes of objects
(see below, also Fan et al.~2000b, Zheng et al.~2000).

 \subsection{Selection  Procedure}

Because of the rarity of high-redshift quasars, and the overwhelming
number of contaminants, our photometric selection procedure of $z>5.8$ 
quasar candidates is quite involved, and includes five separate steps:

1. Selection of $i$-dropout sources from the SDSS database.

2. Running an improved cosmic ray classifier and visually inspecting all
remaining candidates to reduce the number of false candidates due to cosmic rays.

3. Matching the candidate list to the Two-Micron All Sky Survey
(2MASS, \cite{2MASS}) in the public release area of 2MASS, to obtain
$JHK$ photometry.  

4. Independent $z$ photometry to further eliminate false and biased $z$ 
detections.

5. $J$-band photometry of objects not in 2MASS to separate quasar and cool dwarf candidates.

This then leaves us with a sample for follow-up spectroscopy. 

In this and the subsequent section, we describe this process in
detail.  The techniques and rationale for these various cuts are
described in the present section, while the observations themselves
are described in \S~\ref{sec:observations}. 

\subsubsection{Selection of $i$-dropout objects from SDSS imaging data}

The SDSS photometric data are processed by a series of automated
pipelines to carry out astrometric and photometric measurements.
The photometric pipeline ({\it Photo}, \cite{PHOTO}) reduces
the data from the imaging camera and produces corrected images
and object catalogs.
{\it Photo} classifies detected objects either as
``star'' (point source, consistent with the point spread function,
PSF) or ``galaxy'' (extended source).  
However, we select all $i$-dropout objects irrespective of their
star-galaxy classification,  as the star-galaxy separation becomes 
less reliable at low S/N. 
Cosmic rays are recognized as
objects with profile gradients significantly steeper than that of the PSF, 
and are interpolated over; they are not output explicitly by the pipeline.  
While {\it Photo} classifies more than 99.9\% of
the cosmic rays correctly, visual inspection shows that 
a majority of objects with $z$-only detections that are classified 
as ``stars'' still have the appearance of cosmic rays.
This is further confirmed by comparing those areas of the sky with multiple 
observations.
Therefore, we have recently improved the performance of the
object classifier, as follows. 

For every object detected by the imaging pipeline, 
we count the pixels in an $11\times 11$ pixel region
($4.4''\times4.4''$) centered on the object, for which
\begin{eqnarray}
{\hbox{PSF}(d) \over \hbox{PSF}(0)} > {I(d) + c*N[I(d)] \over I(0) - c*N[I(0)]}
\end{eqnarray}
where $I(0)$ is the object's central intensity, $I(d)$ is the
intensity of the pixel in question, PSF(0) and PSF($d$) are the
equivalent quantities for the locally-determined model of the PSF, the
noise $N$ is the photon noise with an additional $0.05 I(0)$ added in
quadrature, and $c = 1.5$.  This criterion searches for objects with
profile gradients significantly sharper than that of the PSF.  If more than 3
pixels satisfy this condition, we consider the object to be a
candidate cosmic ray, and reject it.

The photometric pipeline assigns detailed flags for each detected
object, indicating objects whose photometry (and therefore, colors)
may be in error.   We reject objects flagged as 
being saturated in any band, lying on the bleed trail of
a saturated star, or overlapping the edge of the image boundary.
We pay close attention to objects blended with close neighbors,
and  reject deblended children with {\tt PEAKCENTER}, {\tt NOTCHECKED}, or {\tt
DEBLEND\_NOPEAK} set in the $i$ and $z$ bands, which are typical 
signs of deblending problems.
For a complete description of the definition of flags, see \cite{EDR}.
The flag-checking procedure used here is similar to that used in
the SDSS quasar target selection pipeline (\cite{QTS}).

We have re-classified all the objects with $i^* - z^* > 2.2$
using the new cosmic ray classifier.
We found that depending on the seeing of a specific run, 
$50\%$ to $70\%$ of the $i$-dropout objects formerly 
classified as ``stars'' are now classified as cosmic rays, while
a negligible fraction of the ``real'' objects (based on multiple 
observations and objects with confirmed spectra) are mis-classified 
as ``cosmic rays''.
We then visually inspect the $z$ images of the remaining candidates,
and reject a further 25\% as cosmic rays and other artifacts.

After the improved cosmic ray rejection and visual inspection, we carry out 
independent follow-up $z$ photometry of all the remaining candidates
(of which there are 121; see below) 
using the Apache Point 3.5m and other telescopes (see \S 3.1). This serves two purposes:
(a) to confirm that the objects are indeed real; and 
(b) to measure the $z^*$ magnitude more accurately.
Since our $i$-dropout candidates have the reddest $i^* - z^*$ colors,
we tend to pick up objects lying on the several $\sigma$ tail of the
$z^*$ error distribution (note that the $z$-band only objects are
very faint, and therefore are measured at low S/N); 
%our $z^*$ measurements are
%relatively low S/N).  
%with large positive  as $i$-dropout
%candidates, in the presence of large $z^*$ measurement error, our sample
%would bias towards objects with the lest reliable $z^*$ measurement, i.e.,
%those objects at $>3 \sigma$ tail of the error distribution.
they will thus have artificially red $i^* - z^*$ color and blue $z^* - J$ 
color, and are more likely to be selected as $z>5.8$ quasar candidates 
(see below).
To include those objects in the sample would lower our spectroscopic success
rate considerably.

\subsubsection{Near-IR Photometry: separating quasars and L/T dwarfs}

As discussed by Fan et al~(2000b) and Zheng et al.~(2000), and
illustrated in Figure 1, the major astrophysical contaminants of $i$-dropout
quasars are very cool dwarfs with spectral types L and T
(Strauss et al.~1999, Tsvetanov et al.~2000, Fan et al.~2000a,
Leggett et al.~2000, 2001, Geballe et al.~2001).
These have effective temperatures $T_{eff} < 1800 K$, and 
$i^* -z^* > 2$, and cannot be distinguished from $z>5.8$ quasars
using SDSS optical photometry alone.
Figure 2 presents the $i^* - z^*$ vs. $z^* - J$ color-color diagram
of the $i$-dropout sample.
Finlator et al.~(2000) have matched the SDSS photometric catalog in
$\sim 50$ deg$^2$ with that of 2MASS in the $JHK$ bands.
The small dots in Figure 2 show the SDSS-2MASS matches and 
indicate the location of normal stars on this color-color diagram.
The track of quasar colors shows that by using $z^* - J < 1.5$,
$i$-dropout cool dwarfs and high-redshift quasars are clearly
separated, as the $z^*-J$ color of high-redshift quasars is
dominated by the blue power law continuum, while the cool temperature
of L/T dwarfs results in very red $z^* -J $ color, typically larger than 2.
Therefore we determine whether the candidate has been detected by
2MASS. 
While the non-detection of a faint candidate by 2MASS 
does not give a tight constraint on the spectral shape of the object,
most of the brighter objects ($z^* < 19$) that are located in the
area of the 2MASS publicly released data are indeed detected by 2MASS,
with $z^* -J \gtrsim 2$, indicating that most of them are cool dwarfs.
For those not matched with 2MASS, we carry out follow-up $J$-band 
photometry (see \S3.1).

\subsubsection{Color selection criteria}

The final photometric selection criteria for this survey are:

\begin{equation}
 \begin{array}{l}
 (a)\ z^* < 20.2, \\
 (b)\ \sigma(z) < 0.1, \\
 (c)\ i^* - z^* > 2.2, \\
 (d)\ z^* - J < 1.5.
 \end{array}
\end{equation}
We also require that the object not be detected in any of the other
three bands $u,\ g$ and $r$.

The magnitudes and colors in Eq. (1) are {\em not} corrected for
interstellar extinction.  
This is a small effect, which has been taken into account in the selection 
function calculation (\S5.1). 
In the selection,
only objects with S/N $> 10$ in the $z$ band are considered (item [b]),
as the $i-z$ color becomes very unreliable for the faintest sources
and would have selected many more candidates with a more inclusive cut.
Item (c) selects quasars with $z\gtrsim 5.8$, and
item (d) separates quasars from cool dwarfs. 
Note that the SDSS photometry is on the AB magnitude system 
(Fukugita et al.~1995),
while the $J$-band photometry is on the Vega-based system.
The selected area is illustrated as the shaded region in Figure 2.

\section{Discovery of Three New Quasars}
\label{sec:observations}

 \subsection{Photometric Observations}

 Table 1 lists the details of the 25 SDSS photometric runs used in
this survey. The projection of these runs on the sky is illustrated in
Figure 3. The entire region is located in the Northern Galactic Cap.
In Table 1, the location of each run is presented in terms of
survey stripe number ($n$) and the range of survey great circle longitude
($\lambda_{\rm min}$ and $\lambda_{\rm max}$).
When taking data, the SDSS telescope moves along a series of
great circles on the sky and the photometric camera drift-scans
at the sidereal rate, with an effective exposure time of 54.1 seconds.
The  survey coordinate
system ($\lambda$, $\eta$) is a spherical system with poles at $\alpha_{2000} =
95^\circ$,  and $275^\circ$, $\delta_{2000} = 0^\circ$.
$\lambda = 0^\circ$, $\eta=0^\circ$ is located at $\alpha_{2000} = 185^\circ$, $\delta_{2000} =
35^\circ$ (York et al.~2000, \cite{EDR}). 
Each drift scan tracks a survey stripe $n$ centered on a constant $\eta$, given by
\begin{equation}
\eta = (n-10) \times 2.5^\circ - 32.5.
\end{equation}
Two scans, or strips, one offset to the north and one to the south, are required to fill a stripe.
%For detailed information of the SDSS photometric scan geometry, see \cite{EDR}.
The 25 SDSS  imaging runs used in the $i$-dropout survey in this paper were taken 
between  March 20, 1999 (run 745) and March 19, 2001 (run 2190) over a range of
seeing conditions.
The seeing of SDSS images is characterized by the {\tt psfWidth} parameter,
defined as the effective width of the best fit double-Gaussian PSF model.
For a single Gaussian profile, {\tt psfWidth} = 1.06 FWHM.
Only those runs, or parts of runs,  with {\tt psfWidth} in $z$ band in the
4th column of the SDSS camera (near the middle of the focal plane) better than
$1.8''$ are used in this paper.
The median seeing values (over all six columns of the camera) in the $i$ and $z$ bands are also listed in Table 1.
Due to  the complicated geometry of the survey and overlap between runs, 
we estimated the total survey area by Monte-Carlo integration -- calculating 
the fraction of randomly distributed points on the celestial sphere falling 
into the boundary of any of the runs listed in Table 1.
The total area covered is 1550 deg$^2$, excluding regions with poor seeing. 
The median {\tt psfWidth} is $\sim 1.5''$ in the $i$ and $z$ bands over the whole area.

Following the procedures described in \S2, we first search the SDSS database 
and eliminate false detections (Steps 1 and 2) based on the SDSS images. 
This results in a total of 121 objects with $i^* - z^* > 2.2$ that
require further $z$ and $J$ band photometry.
The candidate list includes five known T dwarfs from
the SDSS (Strauss et al.~1999, Tsvetanov et al.~2000, Leggett et al.~2000),
one known T dwarf from 2MASS (Burgasser et al.~1999),
and the known $z=5.8$ quasar SDSS 1044--0125.
Thirty-five objects are located in the 2MASS first and second
incremental data release area.
Of these, 13 objects are detected in 2MASS $JHK$ photometry and
have colors of L to T dwarfs.

Independent $z$ photometry (Step 4 in \S2.2) was carried out using the 
Seaver Prototype Imaging camera (SPICAM) in the SDSS $z$ filter on 
the ARC 3.5m telescope at the Apache Point Observatory in several nights
between February and April 2001.
SPICAM has a backside illuminated SITe $2048 \times 2048$ pixel CCD  with a
field of view of $4.8'$.  The pixel scale is $0.14''$, and the typical
seeing was $1''$.   The quantum efficiency peaks at 89\% at 6500\AA,
and is about 49\% at 9000\AA. 
The exposure times were between 90 and 180 seconds, depending on the
weather conditions.
Note that since the SDSS photometry itself provides many local
reference stars, photometric weather is not required for $z$ 
photometry.
The S/N is typically twice that of the SDSS imaging.
The SPICAM images are reduced using normal IRAF procedures.
Sky flats are constructed from the scaled median of target images.
$z^*$ magnitudes are measured with aperture photometry with a radius
of $\sim 2''$, and calibrated by local SDSS 
standards. 

Follow-up $J$ band infrared photometry (Step 5 in \S2.2) 
was carried out on photometric nights using several telescopes, between 
February and May 2001.
The majority of the targets were observed with the GRIM II instrument
(the near infrared GRIsm spectrometer and IMager), also on the ARC 3.5m.
It uses a $256 \times 256$ HgCdTe detector and covers a field of view of
$1'$ at f/10.
JHK standards from Persson et al.~(1998) were monitored throughout the night.
We used similar IRAF procedures to reduce the GRIM II data.
A small number of objects were observed with the UFTI instrument, 
a near IR camera with a $1024\times 1024$ HgCdTe detector covering 
$92''\times 92''$ on the 3.8m United Kingdom Infrared Telescope (UKIRT) 
on Mauna Kea,
and with the MAGIC instrument, a near-infrared imager 
and low-resolution infrared spectrograph at the 2.2m telescope
at Calar Alto Observatory in Spain.
The typical photometric error of our infrared photometry is
0.05 -- 0.10 mag, and is dominated by calibration errors.
Since quasars and L/T dwarfs have $z^* - J$ colors that differ by more
than 0.5 mag, this error has very little effect on  
the selection efficiency (\S5.2).

We attempted to follow the sequence described in \S2.2 to select
$i$-dropout quasar candidates. % whenever we can.
But as we had to use photometric nights for the IR photometry,
we changed the observing sequence in two cases: (a) 
on a dry photometric night, IR photometry had the highest priority;
(b) when the weather was non-photometric, 
for objects that had SPICAM $z$ photometry but did not yet have
$J$ band photometry, we carried out spectroscopy directly, without 
pre-selection using $z^* - J$ color. In the latter case, we started 
from either the brightest or the reddest object, depending on weather.
It is worth noting that the rapid instrument change mechanism
at the ARC 3.5m ($\sim$ 15 minutes) and the excellent weather
monitoring system at APO allows a very efficient use of
telescope time.  In the end, we obtained photometry and/or
spectroscopy of our full sample. 

\subsection{Spectroscopic Observations}

Table 2 summarizes the classifications of the $i$-dropout sample.
Using 2MASS matches and follow-up $z$ and $J$ photometry,
we found that among the 121 $i$-dropout candidates,
35 are false detections, most likely cosmic rays that
still passed the first two steps of selection;
71 have $z^* - J > 2$, and are classified as
cool dwarfs. Ten objects are classified as cool dwarfs
based on spectroscopy; so no $J$ photometry was necessary. 
The spectra of these dwarfs were obtained 
using the Double Imaging Spectrograph (DIS) on the
ARC 3.5m between March and May 2001, and
using the Echelle Spectrograph and Imager (ESI) on Keck II
telescope in March 2001.
One of the candidates, SDSSp J081948.97+420930.2,
has an intermediate $z^* - J$ color ($z^* - J = 1.7$);
it is a low-ionization BAL quasar
at $z=2.05$ (see Becker et al.~1997, Hall et al.~2001).
The Keck spectrum of this object is presented in Hall et al. (2001).
Finally, four objects have $i^* - z^* > 2.2$ and $z^* - J < 1.5$, 
including the previously known
$z=5.8$ quasar SDSS 1044--0125, and three new quasars at $z>5.8$.
Note that in this survey, {\em all} objects that satisfy the
color selection criteria (Eq.[1]) indeed turn out to be
$z\gtrsim 5.8$ quasars from their spectra.
The magnitude and colors of these objects are plotted on Figures 1 and 2.

Among the cool dwarfs found in this survey, eleven are T dwarfs.
They form a complete flux-limited T dwarf sample.
Six of these objects were previously known; the optical-IR spectra of
four more of them are presented
in Geballe et al.~(2001).
In a separate paper, we will present the photometry of
L and T dwarfs in the $i$-dropout sample, the observations of
the final T dwarf, and 
the analysis of the surface density and spatial densities
of T dwarfs in the solar neighborhood.

The optical spectra of the three new quasars 
were obtained using the DIS on the ARC 3.5m between March and May 2001.
The instrument and data reduction procedure are described in detail 
by Fan et al.~(1999a).
Figure 4 presents the $z$ band finding charts of the three new quasars,
SDSSp J083643.85+005453.3 ($z=5.82$, SDSS 0836+0054 for brevity),
SDSSp J130608.26+035636.3 (SDSS 1306+0356, $z=5.99$) and
SDSSp J103027.10+052455.0 (SDSS 1030+0524, $z=6.28$). 
Table 3 presents their photometric properties, and
Figure 5 shows the ARC 3.5m discovery spectra of the three objects.
The exposure time of each spectrum is 3600 seconds.
The spectra in Figure 5 have been smoothed to a resolution of $\sim 20$\AA. 
The flux calibration has been adjusted to match the SDSS $z^*$ magnitudes.
The locations of the Ly$\alpha$ and NV emission lines are indicated
in the Figure. The expected positions of Ly$\beta$+OVI and the Lyman Limit are
also shown.

A near-IR spectrum of SDSS 1030+0524
was obtained on the night of 11 June 2001, using the NIRSPEC 
(\cite{NIRSPEC}) instrument on the Keck II telescope, as part of
the Gemini/NIRSPEC service observing program, carried out by
Gemini staff scientists Tom Geballe and Marianne Takamiya.
Observations were taken in grating mode, with the N2 blocking
filter and a $0.76''$ slit, which gives a wavelength coverage
of 1.09 $\mu$m -- 1.29 $\mu$m, and a spectral resolution
of $\sim 1500$.
A total of 10 exposures, each with 300 sec exposure time, were
taken using an ABBA sequence, nodding along the slit. 
The weather was photometric with $\sim 0.7''$ seeing.
HIP 48414, a bright A1V star, was observed as an atmospheric calibration star at
similar airmass to that of the quasar observation.
The NIRSPEC data were reduced using the WMKONSPEC package,
an IRAF package developed at Keck Observatory to reduce NIRSPEC data.
The flux calibration has been adjusted to match its measured $J$-band
magnitude.
The final reduced spectrum, binned to 4\AA$\,$pixel$^{-1}$, is shown in 
Figure 6. 
For comparison, the optical spectrum of SDSS 1030+0524 is also
shown, and the locations of the Ly$\alpha$, NV and CIV emission lines
are indicated. 

\section{Spectral Properties}

The discovery spectra of the three objects show unambiguous
signatures of very high-redshift quasars, similar
to that of SDSS 1044--0125:
strong, broad and asymmetric Ly$\alpha$+NV emission lines,
with sharp discontinuities to the blue side, due to the
onset of very strong Ly$\alpha$ absorption.
The discovery spectra are of relatively low S/N.
Accurate redshift determination  from these spectra are difficult.
The Ly$\alpha$ emission line is severely affected by the Ly$\alpha$
forest absorption, and the NV line is blended with Ly$\alpha$.
The low S/N does not allow us to use weaker lines such as
OI+SiII$\lambda$ 1302 or SiIV+OIV]$\lambda$ 1400 to determine an accurate
redshift.
The next strong emission line in the quasar spectrum is
CIV$\lambda$ 1549, which is  located beyond 1 $\mu$m for
$z>5.5$, and requires near-IR spectroscopy.
Therefore, we use the near IR spectrum to determine the redshift of
SDSS 1030+0524. The redshifts of the other two quasars were
determined using high S/N Keck spectra described in Paper II.  The
uncertainties in all these redshifts are of order 0.02. 
In this section, we first briefly comment on the major spectral signatures 
of each quasar (\S4.1).
We use the emission lines to constrain the metallicity of
the quasar environment (\S4.2).
Then we calculate the average absorption in the Ly$\alpha$ forest
region, based on the discovery spectra, and comment on
the possible detection of the Gunn-Peterson trough in the spectrum of
SDSS 1030+0524 (\S4.3).

\subsection{Notes on Individual Objects}

{\bf SDSSp J083643.85+005453.3} ($z=5.82$).
This quasar has a similar redshift to SDSS 1044--0125. 
The Ly$\alpha$ emission line is very broad and strong.
A separate NV$\lambda$1240 component is tentatively detected.
The rest frame equivalent width (EW) of the Ly$\alpha$+NV line
is $\sim 70$\AA, quite typical of lower-redshift quasars (Fan et al.~2001b).
The Ly$\beta$+OVI emission line is also clearly detected at
$\sim 7000$\AA; this line was also seen in the spectrum of SDSS
1044-0125.  It is evident that there is detectable flux blueward of 
the Ly$\alpha$ emission line. The universe is still highly ionized 
at $z\sim 5.8$.

SDSS 0836+0054 is an extremely luminous object. With $z^* = 18.74$ and 
$J = 17.89$, it has an absolute magnitude $M_{1450} = -27.62$
($\Omega=1$ model, see \S5.1), 
and is the most luminous quasar discovered at $z>4.5$ to date; in particular, 
it is 0.5 mag more luminous than SDSS 1044--1025. 
SDSS 0836+0054 has a radio counterpart in the FIRST radio survey (\cite{FIRST})
at 20$\,$cm. The radio source is unresolved with a total
flux of $1.11 \pm 0.15$ mJy, and the positional match is better than
$1''$. This is the highest-redshift radio loud quasar known,
and the second radio-loud quasar detected at $z>5$
(the other one is SDSSp J091316.56+591921.5, at $z=5.11$, with a 20$\,$cm
flux of 18.1$\,$mJy; see Anderson et al.~2001).

{\bf SDSSp J130608.26+035636.3} ($z=5.99$).
This is the second highest-redshift quasar known to date.
The Ly$\alpha$ emission line is centered at $\sim 8500$\AA, 
and a separate NV component is also detected.
The rest frame equivalent width of Ly$\alpha$+NV is $\sim 60$ \AA.
The Ly$\alpha$ forest is stronger than that in SDSS 0836+0054,
but there is still detectable flux in the Ly$\alpha$ forest region,
immediately blueward of the Ly$\alpha$ emission line.
The apparent absorption at $\sim 9400$\AA\ is not real, but is due to
imperfect telluric absorption correction in this noisy spectrum.
The object has $z^* = 19.47$ and is detected at the 4-$\sigma$
level in the SDSS $i$ band. It is also a luminous object,
with $M_{1450} = -26.93$ in the $\Omega = 1$ model.

{\bf SDSSp J103027.10+052455.0} ($z=6.28$).
This is the highest-redshift quasar known to date.
At $z^* = 20.05$ and $J = 18.87$, it is also the faintest
quasar found in the current survey, and has $M_{1450} = -26.89$ in the 
$\Omega = 1$ model.
It is close to the flux limit of the survey, and is only detected at
10-$\sigma$ in $z$.
The object is very red, and is completely undetected in the 
SDSS $i$ band. 
The discovery spectrum shows very strong Ly$\alpha$+NV emission,
with a rest-frame equivalent width of $\sim 70$\AA, and
a tentative detection of a separate NV component.
The Ly$\alpha$ absorption is strikingly strong, with almost
no detectable flux blueward of the Ly$\alpha$ emission.
A strong CIV emission line is detected in the near IR spectrum
(Figure 6).
A Gaussian fit to the CIV line profile gives
$z_{em} = 6.28 \pm 0.02$, with a rest frame EW = $31.5 \pm 8.6$\AA.
The continuum shape at 9000 \AA\ $< \lambda <$ 13000 \AA\ is
consistent with $f_\nu \propto \nu^{-0.5}$, as indicated by the
dashed line in Figure 6. 

\subsection{Metallicity of $z\sim 6$ Quasars}

Emission line ratios can be used to measure the metallicity of the gas
in the Broad Emission Line Region (BELR).
There is growing evidence from those measurements that BELR have
roughly solar or higher metallicities even out to $z>4$ (e.g., \cite{HF99}).
We use the spectra of our quasars to estimate the quasar metallicity at 
$z\sim6$.

Of various line ratios, the NV $\lambda$1240/CIV $\lambda$1549 ratio
and NV $\lambda$1240/HeII $\lambda$1640 ratio are particularly useful
abundance diagnostics (\cite{HF93}, \cite{HF99}).
Hamann \& Ferland (1993) examined the chemical evolution of
BELR gas by applying spectral synthesis and chemical enrichment
models to the NV/HeII and NV/CIV line ratios.
They found that nominal BELR parameters predict NV/HeII near unity for 
solar abundance, increasing to $\sim 10$ for $Z \sim 10 Z_{\odot}$.
The sensitivity of NV/CIV to metallicity is  due to the fact
that N is a secondary element, with abundance proportional to $Z^2$, 
while C is a primary element with abundance proportional to $Z$.
Their calculations show that for the ISM associated
with a population with solar abundance, NV/CIV $<$ 0.1.
Luminous quasars at $2 < z < 5$ show NV/CIV $\sim 0.1 - 2$,
indicating a metallicity of $Z_{\odot} \lesssim Z \lesssim 10 Z_{\odot}$.

From the Keck/NIRSPEC $J$-band spectrum of SDSS 1030+0524, we find
that the rest frame EW of CIV is 31.5 $\pm$ 8.6 \AA.
We do not detect HeII in emission at $\sim 11900$\AA,
which indicates that the HeII EW is
smaller than $\sim 5$\AA\ at 3-$\sigma$.
The discovery spectra in Figure 5 show tentative detections of NV emission
in all three quasars. The quality of the spectra does not allow an
accurate fit to the Ly$\alpha$+NV line profile in order to derive
the equivalent width of NV. 
%We estimate that in order for the NV emission
%to be seen as a separate component, rather than be completely hidden in
%the wing of the Ly$\alpha$ emission line (EW $\sim$ 50\AA), the rest frame 
%EW of NV in these quasars has to be $\gtrsim 10$\AA.
If we assume the EW of NV to be of the order 10\AA, and a power law 
continuum with $f_\nu \propto \nu^{-0.5}$,
we find the flux ratios NV/CIV $\gtrsim$ 0.4, and NV/HeII $\gtrsim 3.0$.
Both limits would imply super-solar metallicity of the BELR region.
In the calculation of Hamann \& Ferland (1993), these line ratios
are consistent with the ``Giant Elliptical'' model 
with fast stellar evolution and a top-heavy initial mass function, 
with $3Z_{\odot} < Z < 10 Z_{\odot}$.
A high resolution, high S/N spectrum is needed to fit the
Ly$\alpha$+NV line profile and put better constraints on the
line ratios (Paper II).
However,  note that Krolik \& Voit (1998) point out that NV can be excited by 
resonance scattering from the red wing of the Ly$\alpha$ emission
line, thus biasing these metallicity calculations.

\subsection{Average absorption in the Ly$\alpha$ forest region}

The luminous high-redshift quasars discussed in this paper
are ideal targets for detailed studies of the intergalactic medium
at high redshift.
%In Paper II, we use high S/N, high resolution Keck and VLT spectra
%to investigate the properties of Ly$\alpha$ absorptions in these
%quasars, in particular the nature of the absorption in the highest
%redshift object, SDSS 1030+0524.
We calculate  average absorption in the 
Ly$\alpha$ forest region based on the discovery spectra.
The results are summarized in Table 4, which also includes the
measurements for SDSS 1044--0125.
Three quantities are calculated for each quasar.
We first estimate the average continuum decrements as:
$D_{A,B} \equiv \left\langle 1 - f_\nu^{obs}/f_\nu^{con} \right\rangle $, where
$f_\nu^{obs}$ and $f_\nu^{con}$ are the observed and the unabsorbed
continuum fluxes of the quasar, and
$D_{A}$ and $D_{B}$ measure the decrements in the region between
rest-frame Ly$\alpha$ and Ly$\beta$ ($\lambda = 1050 - 1170$ \AA) and
between Ly$\beta$ and the Lyman Limit ($\lambda = 920 - 1050$ \AA),
respectively (\cite{OK82}).
Following Fan et al.~(2000b), we measure $D_{A}$ and $D_{B}$
assuming a power law continuum $\nu^{\alpha}$ with $\alpha = -0.5$.
The errors on $D_A$ and $D_B$ are dominated by the continuum determination.
The error bars in Table 4 reflect the range of $D_A$ and $D_B$,
allowing $\alpha$ to range from $-1.5$ to +0.5.

The range of IGM redshifts covered by the window over which 
$D_{A}$ is calculated is rather large ($5.0 < z_{abs} < 5.7$ for
a quasar at $z=6.0$).
Therefore we also calculate the transmitted flux ratio over a smaller
window (of 0.2) centered on a redshift $z_{abs} = z_{em} - 0.3$:
\begin{equation} 
{\cal T}(z_{abs}) \equiv \left\langle f_\nu^{obs}/f_\nu^{con} \right\rangle,
\hspace{1cm} (1+z_{abs}-0.1)\times 1216 \hbox{\AA} < \lambda <
(1+z_{abs}+0.1) \times 1216 \hbox{\AA}.
\end{equation}
%At a given $z_{abs}$, it is the percentage of unabsorbed flux remaining
%in the wavelength window that covers only $z_{abs} \pm 0.1$.
We assume a power law continuum of $f_\nu \propto \nu^{-0.5}$.
The error bars in Table 4 reflect only the photon noise of the spectra.
The quantity $\mathcal{T}$ is more noisy than $D_A$, but can be
calculated up to within a redshift of 0.2 of the quasar; closer than
this, it would be affected both by the proximity effect from the
quasar itself and the blue wing of the Ly$\alpha$ emission
line ($\Delta z \sim 0.2$,  Oke \& Korycansky 1982).
In Table 4 and Figure 7, we show the $\mathcal{T}$$(z_{abs})$ values for
the redshifts $z_{abs} \sim z_{em} - 0.3$. 

From Table 4 and Figure 7, it is evident that the amount of 
Ly$\alpha$ absorption increases rapidly with redshift,
with $D_{A} > 0.9$ at $z_{em} > 5.8$.
For comparison, $D_{A} \sim 0.75$ at $z_{em} \sim 5$ 
(Songaila et al.~1999).
The most striking feature, however, is that based on the
spectrum of SDSS 1030+0524, $\mathcal{T}$$(z_{abs} = 6.0) = 0.003 \pm 0.020$,
consistent with no flux detected at all.
The flux level is consistent with zero from 8400\AA\ to
8700\AA\ in the spectrum (Figure 5), corresponding to $5.9 < z_{abs} < 6.15$.
The 1-$\sigma$ {\em lower limit} on the flux decrement 
in this quasar is a factor of 50. 
At slightly lower redshift, we find $f/f_{\rm con} (z_{abs} = 5.5 - 5.7) =
0.07 - 0.10$ from the other three $z>5.8$ quasars in the sample.
The flux decrement observed in SDSS 1030+0524
is at least a factor of 4 larger than that seen in any other quasar at $z>5$. 
There seems to be a drastic change in the amount of Ly$\alpha$
absorption at $z\sim 6$.

\subsection{Gunn-Peterson Effect in the Spectrum of SDSS 1030+0524}

The absence of flux over a 300\AA\ region in the ARC 3.5m
discovery spectrum  indicates a possible
first detection of the complete Gunn-Peterson trough.
This detection is still highly tentative, due to the
uncertainty in sky subtraction in these low S/N spectra.
Note that spectra taken on different nights show the
same non-detection of flux in the Ly$\alpha$ forest region.
This result shows that the fraction of neutral hydrogen has increased
substantially between $z=5.7$ and $z=6$, and, if confirmed
by high S/N spectroscopy, would show that the universe is 
approaching the epoch of reionization.
In Paper II, we use the high-resolution Keck spectrum of
SDSS 1030+0524 to place stronger constraints on the status of the IGM
at $z\sim 6$, and discuss its implication on the reionization of the
universe in more detail.

For a uniformly distributed IGM, the Gunn-Peterson (1965) optical depth is:
\begin{equation}
\tau_{GP} (z) = 1.8 \times 10^5 h^{-1} \Omega_M^{-1/2} 
\left( \frac{\Omega_b h^2}{0.02} \right)
\left ( \frac{1+z}{7} \right )^{3/2}
\left( \frac{n_{\rm HI}}{n_{\rm H}} \right ). 
\end{equation}
If the IGM were mostly neutral, $\frac{n_{\rm HI}}{n_{\rm H}} \sim 1$,
and the Gunn-Peterson opacity would be $\sim 10^5$.
Even a tiny fraction of neutral hydrogen (of the order of $10^{-5}$)
in the IGM could result in a large optical depth, and undetectable flux 
in the Ly$\alpha$ forest region.
%In reality, the IGM is highly non-uniform. The reionization of the
%universe happens in a complicated, but most likely gradual fashion
%(\cite{MHR00}).
Therefore, the existence of the Gunn-Peterson trough by itself does not 
indicate that the object is observed prior to the reionization epoch, as it
takes only a tiny fractional abundance of neutral hydrogen to create a
trough.  Indeed, both semi-analytic models (Miralda-Escud\'{e} et al. 2000)
and hydro-dynamical simulations (\cite{Gnedin00}) of structure formation 
show that the Gunn-Peterson trough should begin to appear at $z\sim6$, 
while the epoch of overlap of HII regions could have occurred at a higher 
redshift.

\section{Spatial Density of $z\sim 6$ Quasars}

\subsection{A Color-selected Complete Sample at $z>5.8$}

The three new $z>5.8$ quasars presented in this paper, plus
SDSS 1044--0125, comprise a complete color-selected flux-limited sample
of $z>5.8$ quasars at $z^* < 20.2$.
The color selection criteria are listed in Eq. (1), and the
total sky coverage is 1550 deg$^2$, using the SDSS photometric
runs in Table 1.
The continuum properties of this sample are given in Table 5.
Following Fan et al.~(2000c), the quantity
$AB_{1280}$ is defined as the AB magnitude of the continuum 
at rest-frame 1280\AA, after correcting for interstellar 
extinction using the map of \cite{Schlegel98}.
We extrapolate the continuum to rest-frame 1450\AA,
assuming a continuum shape $f_\nu \propto \nu^{-0.5}$ to 
calculate $AB_{1450}$. 
%The rest frame 1450\AA, at $> 1$ micron at
%$z>5.8$, is beyond the spectral range of optical spectrum.
Table 5 also lists the absolute magnitudes
$M_{1450}$ and $M_{1280}$ in both the $\Lambda$ and 
$\Omega=1$ models.
In the next subsections, we first calculate the selection function
of our color-selected sample, then derive the total spatial density
of luminous quasars at $z\sim 6$.

\subsection{Selection Function}

The selection function $p(M_{1450},z)$ is defined as the 
probability that a quasar of a given $M_{1450}$ and $z$ will
satisfy the selection criteria (eq. [1]).
We calculate this using a Monte-Carlo simulation of quasar colors,
based on the quasar spectral model described in Fan (1999).
We follow the procedures described by Fan et al.~(2001a),
with several modifications:

(1) $J$ band magnitudes are included in the calculation.
We use the $J$ band filter curve measured for the filters used
in the GRIM II instrument on the ARC 3.5m (we assume that the 
the UFTI and MAGIC instruments have similar $J$ band filter curves).
It is very similar to the filter used in the 2MASS survey. The detailed filter 
curve is available from the authors. 
Since the $J$ band flux of the quasar is dominated by the power law
continuum, any small differences between the $J$ band filter curves of
different instruments has
little effect on the selection function.

(2) SDSS photometric errors, as a function of seeing, are modeled
in more detail. The selection criteria include $\sigma(z^*) < 0.10$,
which is a strong function of the $z^*$ magnitude and the observed
seeing. 
We first construct the distribution of $i$ and $z$ seeing for
the entire survey area. 
In the simulation, we randomly draw the $i$ and $z$ seeings from
this distribution and use them to calculate the photometric error terms
of the SDSS magnitudes.

(3) The selection is based on the $z^*$ magnitudes without correcting
for interstellar extinction. We therefore construct the 
$E(B-V)$ distribution of the entire survey area based on the dust map of 
\cite{Schlegel98},  and assign values of $E(B-V)$ drawn at random from 
this distribution, to the quasars in the simulation.

Following Fan et al.~(2001a), we model the quasar intrinsic 
spectral energy distributions (SEDs) using a power law continuum
$f_\nu \propto \nu^{-\alpha}$, with $\alpha = 0.79 \pm 0.34$, 
plus a series of broad emission lines, with the rest-frame equivalent width
of Ly$\alpha$+NV with a mean of 69.3\AA\ and a standard deviation of 18.0\AA.
We use the quasar absorption line models in Fan (1999) to simulate
the Ly$\alpha$ absorption, which is critical to predicting
the colors of high-redshift quasars. 
In this model, we assume that the number density
of Ly$\alpha$ forest lines evolves as $N(z) \propto (1+z)^{2.3}$.

The results of the selection function are summarized in Figure 8
for both cosmologies. The heavy lines in the figure represent the
5\% contour, indicating the survey limit in absolute magnitude at
each redshift.
From the figure, we notice:

(1) The survey is sensitive to the redshift range 
$5.8 \lesssim z \lesssim 6.3$. The lower limit is due to 
the cut $i^* - z^* > 2.2$. The upper limit is due to both the
cut $z^* - J > 1.5$, and the fact that quasars begin to disappear
from $z$ band images because of Ly$\alpha$ absorption in the $z$ band (\S2.1).

(2) At $5.8 \lesssim z \lesssim 6.0$, the survey reaches $M_{1450} \sim -26.5$, 
and is rather complete ($p > 0.8$) at the bright end.
At $z>6.0$, it becomes increasingly incomplete even for $M_{1450} < -27$.
The survey limiting absolute magnitude as a function of
redshift changes much faster than the change in distance modulus would
indicate, due to increasing Ly$\alpha$ absorption in the $z$ band.

Figure 8 also plots the locations of the four quasars in the
sample. Three of them are located in the region of high selection
probability $(p > 0.6)$, while the highest-redshift and faintest object, 
SDSS 1030+0524, has a selection probability of only $\sim 20\%$.

These calculations are based on the statistics of the quasar emission line,
continuum and absorption systems observed at much lower redshift, $z\sim4$. 
To test the sensitivity of our conclusions to our
imperfect knowledge of quasar SEDs at $z\sim6$, we repeated the calculations
with various continuum slopes and emission line strengths.
%There is no reason to expect these properties to remain the same at
%$z\sim6$. Our ability to simulate quasar colors is limited by our
%lack of knowledge of quasar SEDs at the highest redshift.
%But given the size of the sample, the error on the spatial density
%is dominated by small number statistics rather than the 
%assumptions we have made in the quasar color simulation.
If we assume a different slope for the power-law continuum,
$\langle \alpha \rangle = 0.3$
rather than 0.79, we find that the selection probabilities of the quasars
change by an average of 3\%; similarly, when we change the average
equivalent width of the Ly$\alpha$+NV blend from 69\AA\ to 39\AA, the
selection probabilities change by 2\%.  Thus, given the small size of the
quasar sample, the error in the estimated spatial density 
(\S5.3 below) is dominated by small number
statistics rather than the assumptions made in the quasar color simulation.
However, using these simulations we find that the selection criteria in 
Eq. (1) will not select quasars with undetectable emission lines 
(EW(Ly$\alpha$) $\sim$ 0, e.g., SDSS 1533-0039, \cite{BLLAC}). 
Such objects will have bluer $i-z$ and redder $z-J$ colors, and are located
much closer to the stellar locus. 
They can, however, be selected using a more relaxed color cut. 
e.g., using $i-z>2.0$ and $z-J < 1.8$, the average selection probability
of these unusual quasars will be higher than 40\% for bright quasars 
at $z<6.2$. 

The $i^* - z^*$ color of $z \sim 6$ quasars is mainly determined by the 
Ly$\alpha$ absorption. We have very little idea of how the 
Ly$\alpha$ forest evolves with redshift at $z\sim 6$.
However, as shown in \S4.3, 
there is tentative evidence that the evolution of average absorption
is stronger than if we assume a simple extrapolation from lower redshift.
If that is the case, then at a given redshift, we will overestimate
the $i$ flux in our models, while the $z$ flux remains almost unchanged, 
as it is dominated by the quasar intrinsic spectrum (continuum + 
emission lines) redward of Ly$\alpha$.
Thus the true $i^* - z^*$ color of quasars is redder than the model we 
assume here, while the $z^* - J$ color changes little.
This means that our $i^*-z^*$ color cut would be sensitive to quasars
at redshifts slightly below 5.8; the selection probability at $z>5.8$
would remain unchanged. 

The presence of foreground dust in the quasar environment would redden
the quasar colors. In Figure 2, we indicate the reddening vector
for $E(B-V) = 0.10$ for a quasar at $z=6$. 
As discussed in Fan et al.~(2001b), the number
density of high-redshift quasars could be significantly affected by
the uncertain amount of dust extinction.

\subsection{Spatial Density of Luminous Quasars at $z\sim 6$}

With four quasars in the sample, we can only derive the total
spatial density of the quasars in the redshift and luminosity
range the survey covers.
We calculate this quantity using the $1/V_a$ method, following
the discussion in Fan et al.~(2001a). For each quasar, the
volume over which it would have been observed in our survey
is
\begin{equation}
V_a = \int_{\Delta z} p(M_{1450}, z) \frac{dV}{dz} dz,
\end{equation}
where the integral extends over the redshift range $5.7 < z < 6.6$.
The total spatial density and its statistical uncertainty can be
estimated as:
\begin{equation}
\rho = \sum_i \frac{1}{V_a^i}, \hspace{0.5cm} 
\sigma(\rho) = \left [ \sum_i \left ( \frac{1}{V_a^i} \right )^2 \right ]^{1/2}.
\end{equation}

Using the selection function presented in \S5.1, we find that
at the average redshift of $\langle z \rangle = 5.97$,
$\rho (M < -26.8) = (1.14 \pm 0.58) \times 10^{-9}$ Mpc$^{-3}$
in the $\Omega=1$ model,
and $\rho (M < -27.1) = (0.70 \pm 0.35) \times 10^{-9}$ Mpc$^{-3}$
in the $\Lambda$-model.

Schmidt, Schneider \& Gunn (1995, hereafter SSG) derive the high-redshift 
quasar  luminosity function in the range $2.7 < z < 4.75$ using 90 quasars at 
$M_{1450} \lesssim -26$.
Fan et al.~(2001b) calculate the evolution of the quasar luminosity
function over the range $3.6 < z < 5.0$ and $-27.5 < M_{1450} < -25.5$,
using a sample of 39 quasars; the two luminosity functions are in 
good agreement. 
In Figure 9, we show the density of quasars at $M_{1450} < -26.8$ 
found in this paper, along  with the results from SSG, Fan et al.~(2001b),
and the 2dF survey (Boyle et al.~2000) at $z<2.5$.
The quasar density  at $z\sim 6$ found in this paper is consistent with 
extrapolating the best-fit quasar luminosity functions
in the range  $3 \lesssim z \lesssim 5$ 
from SSG and Fan et al.~(2001b). 
Both SSG and Fan et al.~(2001b) assume a single power law luminosity function
at the bright end and that the number density of quasars declines
exponentially as a function of redshift.
For example, in the $\Omega=1$ model, the maximum likelihood results 
of Fan et al.~(2001b) find that at $z=5$, $\rho(M_{1450} < -26.8) = 
1.98 \times 10^{-9}$ Mpc$^{-3}$. Extrapolating the result to 
$z=6$, we predict $\rho(M_{1450} < -26.8) = 0.65 \times 10^{-9}$ Mpc$^{-3}$,
within 1-$\sigma$ of the spatial density estimated in this paper.
From $z=5$ to $z=6$, the density of luminous quasars drops by a
factor of $\sim 1.8$. 
This drop is consistent with a decline of a factor of 3 per unit redshift
found in Fan et al.~(2001b) and 2.7 per unit redshift found in SSG.

\section{Discussion}

At $M_{1450} \sim -27$, the luminous quasars described in this
paper are likely to reside in the most massive systems at high redshift.
We estimate the masses of the central black holes in these quasars following
the assumptions of Fan et al.~(2000c):
the quasars are emitting at the Eddington luminosity with SED modeled 
by Elvis et al.~(1994), and the fluxes of the observed quasars are not
significantly magnified by gravitational lensing or beaming. 
We find that in the $\Lambda$-model, the estimated black hole masses are
$M_{BH} = 4.8\times10^9 \rm M_{\odot}$, $2.0\times10^9 \rm M_{\odot}$, 
and $1.9\times10^9 \rm M_{\odot}$ for SDSS 0836+0054, SDSS 1306+0356 and
SDSS 1030+0524, respectively.
Under the same assumptions, 
$M_{BH} = 3.4 \times10^9 \rm M_{\odot}$ for SDSS 1044--0125.
In the $\Omega=1$ model, the estimated black hole masses are about 20\% lower.
Note that the Elvis et al.~(1994) quasar SED, which we use to estimate the
(substantial) bolometric correction from the flux at rest wavelength
of 1450\AA, 
is based on quasars at
$z\lesssim 2$. The bolometric correction of quasars at $z\sim 6$ could
in fact be quite different, but to determine this will require
observations of these high-redshift quasars over a large range of wavelengths. 

There is no established relation between black hole mass and the mass of
the galactic bulge in which it resides, at high redshift.  If we make the 
rather large assumption that the Magorrian et al.~(1998) relation,
$M_{BH}/M_{bulge} = 3 \times 10^{-3}$, determined at $z \sim 0$, 
also holds at high redshift,  we find that in the $\Lambda$-model,
$M_{bulge} = (6 - 16) \times 10^{11} \rm M_{\odot}$ for the quasars
in this paper. 
Assuming further that the ratio of the bulge to dark matter halo mass
is of order the ratio of $\Omega_b$ to $\Omega_M$, we find that these
quasars reside in dark matter halos with mass 
$M_{halo} \sim 10^{13}  \rm M_{\odot}$ for $\Omega_b/\Omega_M \gtrsim 10$.
However, Kauffmann \& Haehnelt (2000) argue that the ratio of 
black hole mass to bulge mass could be smaller at high-redshift
(see also \cite{Ridgway01}).
Alternatively, using the Gebhardt et al.~(2000) $z \sim 0$ relation
between black hole mass and bulge velocity dispersion 
$M_{BH} = 1.2 \times 10^8 \rm M_{\odot} (\sigma_{\rm bulge}/200 \rm \ km\ \rm
s^{-1})^{3.75}$,
we find $\sigma_{\rm bulge} = (420 - 530)\, {\rm km\, s^{-1}}$.

Note that we assume that these four high redshift quasars are not magnified
by gravitational lensing. In a $\Omega = 1$ universe, the lensing 
probability rises steeply with redshift up to $z \sim 1$, then
becomes rather flat.
In a $\Lambda$-dominated model, the increase in lensing
probability stays steep to somewhat higher redshift, but flattens
above the redshift at which the universe is matter-dominated. 
Using the equations of Turner (1990), we find the raw lensing probability
to be $\sim 0.01$ at $z\sim 6$ for reasonable assumptions for the  lensing
optical depth.
The probability of magnification bias, however, could be much larger
since the sample in this paper is a flux-limited sample, and the observed
quasars are on the tail of a steep luminosity distribution
(Blandford \& Narayan 1992).
A detailed calculation of magnification bias is beyond the scope
of this paper.

Following Turner (1991) and Haiman \& Loeb (2001), we  consider the growth 
of supermassive black holes in high-redshift quasars.
The $e$-folding timescale for black hole growth is
$4 \times 10^7 (\frac{\epsilon}{0.1}) \eta^{-1}$ yr,
where $\eta^{-1}$ is the ratio of the quasar bolometric luminosity to 
the Eddington luminosity, and  $\epsilon$ is the radiation efficiency.
For SDSS 0836+0054, the most luminous quasar in our sample,
it takes 20$\eta$ $e$-folding times, or $\sim 0.8$ Gyr,
to grow from a 10 $\rm M_{\odot}$ stellar black hole, if $\epsilon =1$.
This timescale is very close to the age of the universe 
at $z\sim 6$ in the $\Lambda$-model.
If $\eta = 1$, it is actually {\em longer} than the age
of the universe in the $\Omega=1$ model.
Note that the black hole could grow much faster if
the radiation efficiency is very low.
Alternatively, the black hole could have formed from
a much more massive seed black hole, such as a very massive
object in Population III (VMOs; Bond, Arnett \& Carr 1984),
or from mergers of smaller black holes (e.g. Kauffmann \& Haehnelt
1999; Menou, Haiman, \& Narayanan 2001). 

\subsection{Constraint on the Shape of the Quasar Luminosity Function}

The quasar luminosity function at high redshift is a sensitive test
of cosmological parameters and models of quasar evolution.
The observed luminous quasars most probably represent rare peaks in the 
density field at $z \sim 6$, and hence probe the exponential, high-mass 
tail of the underlying dark matter halo distribution.
The slope of the luminosity function is determined
by both the slope of the halo mass function and the relation between black 
hole mass (proportional to quasar luminosity if the quasar is radiating at the 
Eddington limit) and dark halo mass.
The $z\sim6$ quasars presented in this paper are extremely rare
objects, representing many-$\sigma$ peaks in the density field.
One would therefore expect their luminosity function to be very steep.
We use the luminosity distribution of the four quasars in our sample to
constrain the slope of the bright end of the quasar luminosity function, and
compare it with theoretical expectations, assuming that the fluxes
of the four observed quasars are not significantly magnified by
gravitational lensing.

Among the four quasars in the complete sample in Table 5, 
two have $M_{1450} < -27$ ($\Omega=1$ model).
We use this as a constraint to calculate the expected number of fainter
quasars in the survey area, assuming a power law luminosity function, 
$\Psi(L) \propto L^{\beta}$, and taking into account the survey selection 
function presented in Figure 8.  We then compare this expected number with the
actual number of lower-luminosity quasars observed
(two at $M_{1450} > -27$).
For $\beta  >  -3.9$, we expect to find less than six quasars at  
$M > -27$ in our survey.
In this case,  the probability of observing 
fewer than three lower luminosity quasars is larger than 5\% 
(assuming Poisson statistics).
Therefore, $\beta  >  -3.9$  is consistent with the observed distribution
at greater  than the 2-$\sigma$ level.
Note that this constraint encompasses the values of the 
bright-end slope of the quasar luminosity function
from Fan et al.~(2001b),
who found $\beta = -2.6$ using a sample of quasars at $z\sim4$,
%For this luminosity function, we expect to find 1.95 quasars at
%$ M > -27$ in our survey, in good agreement with the two observed. 
%Using the steeper luminosity function observed at $z<2$,
and Boyle et al. (2000), who found $\beta = -3.4$ at $z<2$.
%low-luminosity quasars 
%in our survey. This slope is still consistent with the observed
%number of lower luminosity quasars, as the probability of finding less 
%than three quasars in the survey is 0.34, assuming Poisson statistics.
However, the observed luminosity distribution is not consistent with an even
steeper luminosity function. 
The probability drops to 0.01 for $\beta = -4.3$, and to 0.001 for $\beta = -4.6$.
In other words, a steep quasar luminosity function with
$L  \propto L^{-4.6}$ is ruled out at the $\sim 3$-$\sigma$ level by
the current observations.
This constraint simply reflects the fact that even after taking into
account the lower selection completeness of the survey at the faint end,
there are not many  lower luminosity quasars in our survey at $z\sim6$
(only two are observed), given 
that we have discovered two bright ones ($M_{1450} < -27$, $z^* < 19.3$).
We note that this constraint on the slope of the bright end of the
luminosity function is quite weak, as it is based on only four objects, and depends
strongly on our model of the survey selection function 
and the assumption that the observed quasars
are not significantly magnified by gravitational lensing.
In order to obtain a stronger constraint on the slope of the
luminosity function, a large, and more importantly, a
deeper sample of $z>6$ quasars is needed.
%We note that this constraint on the slope of the bright end of the
%luminosity function depends only on the assumptions that our model of
%the survey selection function is reasonably correct, and that the quasars
%have not been significantly magnified by gravitational lensing.

Is this constraint on the bright end slope of the quasar luminosity function 
consistent with a simple model of quasar evolution?  
We can answer this question in the context of specific models of
quasar formation.
Following Haiman \& Loeb (1998), we assume that there is a monotonic relation
between quasar luminosity at high redshift and the mass of the dark
matter halo in which it resides, and assume further that every black hole
shines as a quasar for a given fraction of the age of the universe to
that redshift (the ``lifetime'' or ``duty cycle'' of the quasar).  
We can then match the observed quasar number density at $z=6$ to the
number density of dark matter halos found in the
Hubble-volume simulations (Jenkins et al.~2000), 
if we assume that the quasars in our sample are associated with dark matter 
halos more massive than $M_{\rm halo} = 1-3 \times 10^{13}$M$_{\odot}$.  
In this case, 
the dark halo mass function at this mass scale has a logarithmic slope 
between $-5$ and $-6$, depending on the cosmology and quasar life time
($\sim 10^7 - 10^8$ years).
%Here, we carry out calculations similar to those described in
%Haiman \& Loeb (2001).
%For a given quasar lifetime $t_{\rm QSO}$, the duty cycle
%of quasar phenomena $f_{\rm duty} = t_{\rm QSO} / t_z$,
%where $t_z$ is the age of the universe at redshift $z$.
%Therefore $\rho (>L) / f_{\rm duty}$ gives the intrinsic 
%abundance of dark halos in which the quasar resides.
%If we assume that quasar luminosity is a monotonic function
%of the dark halo mass, for a given cosmology, we can then
%calculate the minimum halo mass corresponding to the survey limiting 
%luminosity, as well as the local slope of the mass function at the 
%relevant halo mass,
%$\rho(M) \propto M^{\gamma}$. We use the mass function found
%in the Hubble-volume simulations (Jenkins et al.~2001).
%
%Using the density of $z\sim6$ quasars found in \S5.2, we
%calculate the minimum halo mass and the slope of mass function
%$\gamma$, for different quasar lifetimes and two cosmologies
%used in this paper.
%In the $\Omega=1$ model, we find that for $t_{\rm QSO} = 4 \times
%10^7$ yr, $M_{min} = 2.9 \times 10^{13} \rm M_{\odot}$ and
%$\gamma = -5.9$.
%For $t_{\rm QSO} = 4 \times
%10^8$ yr, we find $M_{min} = 3.1 \times 10^{13} \rm M_{\odot}$ and
%$\gamma = -6.1$.
%For comparison, in the $\Lambda$-model and for
%$t_{\rm QSO} = 4 \times
%10^8$ yr, we find $M_{min} = 9.0 \times 10^{12} \rm M_{\odot}$ and
%$\gamma = -5.1$.
Note that this minimum halo mass is comparable to that
estimated using the $z\sim0$ relation of Magorrian et al.~(1998),
assuming that the quasars radiate at the Eddington limit.

These luminous quasars represent about 5-$\sigma$ peaks in the 
density field for a $\Lambda$-CDM power spectrum, 
and the mass function is extremely steep at this mass.  
The slope of the dark halo mass function 
%$\rho(M) \propto M^{\gamma}$,
%with $\gamma = -5 \sim -6$ (depending on the assumed cosmology),
is steeper than that of the quasar luminosity function
$\beta > -4.6$.
This difference in the slopes of the bright end of the quasar luminosity
function and the dark halo mass function at the corresponding mass scale
implies that for high-redshift quasars, the luminosity $L$ cannot simply
scale linearly with $M_{\rm halo}$; rather, high-mass black holes radiate
more efficiently than do lower mass ones.
Alternatively, the quasar luminosity function can be shallower than the
mass function if there is large scatter in the $L - M $ relation,
or if the fluxes of the observed quasars have been significantly amplified
by gravitational lensing.
Clearly, the above results are oversimplified and are based on a
number of assumptions regarding both quasar activity and 
the correspondence between quasar luminosity and the mass of the
dark matter halo in which it resides.
However, they illustrate the importance of the measurement of
the shape of the quasar luminosity function in constraining models of 
quasar formation and evolution.

\subsection{Quasar Contribution to the Ionizing Background at $z\sim 6$}

The absence of the Gunn-Peterson trough in SDSS 1306+0356 ($z=5.99$)
indicates that the universe is still highly ionized at $z\sim 5.7$,
presumably by the ionizing photons from quasars and star-forming
galaxies.  The photon emissivity per unit comoving volume required to keep the universe ionized at
$z=6$ in the $\Omega = 1$ model
is (\cite{MHR99}),
\begin{equation}
\dot{\cal N}_{ion}(z) = 10^{51.2}\,{\rm s^{-1}\, Mpc^{-3}}\left({C\over 30}\right)\times
\left(\frac{1+z}{6}\right)^3 \left({\Omega_b h^2\over 0.02}\right)^2,
\end{equation}
where $C$ is the clumping factor of the IGM. 
Based on the extrapolation of the quasar luminosity function
from $z<4.5$, Madau et al.~(1999) conclude that quasars alone do not
provide enough ionizing photons to keep the universe ionized at $z>5$, 
and that the major contribution to the cosmic ionizing background has 
to be produced by star-forming galaxies (see also \cite{SPA01}).
Here we calculate the total contribution of the quasar population to
the ionizing background at $z\sim6$, using the quasar density derived
in \S5.2 and the constraint on the bright end slope of the quasar luminosity 
function in \S6.1.

We calculate the total emissivity per unit comoving volume of ionizing photons from the quasar population:
\begin{equation}
\dot{\cal N}_{Q(z)} = \frac{\epsilon_Q^{1450}(z)}{n_I^{1450}}, 
\end{equation}
where $\epsilon_Q^{1450}(z)$ is the total quasar emissivity at 1450\AA,
with units of erg s$^{-1}$ Hz$^{-1}$ Mpc$^{-3}$,  
assuming a luminosity function $\Psi(L,z)$
\begin{equation}
\epsilon_Q^{1450}(z) = \int \Psi(L_{1450},z) L_{1450} dL_{1450},
\end{equation}
and $n_I^{1450}$ is the total number of ionizing photons for a source with
a luminosity of 1 erg s$^{-1}$ Hz$^{-1}$ at 1450\AA.
We assume the quasar SED following Madau, Haardt, \& Rees (1999).
\begin{equation}
L(\nu) \propto \left\{ \begin{array}{ll}
		 \nu^{-0.3} & \mbox{2500\AA $< \lambda <$ 4400\AA;} \\ 
		 \nu^{-0.8} & \mbox{1050\AA $< \lambda <$ 2500\AA;} \\ 
		 \nu^{-1.8} & \mbox{ $ \lambda <$ 1050\AA,}   

			\end{array}
 \right.
\end{equation}
and integrate
over the energy range from 1 to 4 Rydberg, with a cutoff at 4 Rydberg because 
of HeII absorption. 

We assume that the luminosity function of quasars at $z\sim6$ is described by
a double power law, and write it as a function of absolute magnitude 
$M_{1450}$:
\begin{equation}
\Psi(M_{1450}) = \frac{\Psi^{\ast}}{10^{0.4(\beta_1+1)(M_{1450}-M^\ast_{1450})}+10^{0.4(\beta_2+1)(M_{1450}-M_{1450}^\ast)}},
\end{equation}
where $M_{1450}^\ast$ is the characteristic luminosity.

None of the current quasar surveys at $z>4$ constrains the faint end slope
of the quasar luminosity function. 
For example, Fan et al.~(2001b) only compute the luminosity function at 
$M_{1450} < -25.5$.
Therefore, the characteristic luminosity  $M_{1450}^\ast$  and the 
faint end slope
$\beta_2$ are both unknown at high redshift, and we simply {\em assume}
%$\beta_2$ needs to be smaller than 2 for the integral to converge.
the faint end slope of the luminosity function 
to be the same as determined by Boyle et al.~(2000) at $z<2$, 
$\beta_2 = -1.58$.
Thus, for a given bright end slope $\beta_1$, we can derive $\Psi^\ast$ 
by requiring 
$\int_{-\infty}^{-26.8} \Psi(M) dM = 1.1 \times 10^{-9}$ Mpc$^{-3}$ ,
as determined in \S5.2.

In \S6.1, we demonstrated that the bright end slope $\beta_1$ of the quasar 
luminosity function at $z\sim6$ is consistent with that measured at lower 
redshift ($\beta_1 = -2.58$, Fan et al.~2001b, and $\beta_1 = -3.43$, 
Boyle et al.~2000), and that a much steeper slope can be ruled out 
($\beta_1 < -4.3$ is ruled out at the 99\% confidence level, and 
$\beta_1 < -4.6$ with 99.9\% confidence).
In Figure 10, we calculate the quasar ionizing photon density
integrated over the full luminosity function as a function
of $M_{1450}^\ast$ for a series of values of $\beta_1$, assuming $\beta_2 = -1.58$, and normalizing to the quasar density at $M_{1450} < -26.8$.
The quasar ionizing photon density is compared with what is required to
keep the universe ionized at $z\sim6$ (heavy solid line).

From Figure 10, it is evident that with $\beta_1 > -3.5$, quasars
would not produce enough photons to keep the universe ionized at
$z\sim6$, unless $M_{1450}^\ast > -23$, i.e., only 
if the ionizing background is dominated by low luminosity AGNs can the
quasar luminosity function provide enough ionizing photons.	
This would imply a large negative luminosity evolution of
the luminosity function.
%This would imply a different luminosity function shape from that observed
%at low redshifts.
The universe could be ionized with luminous quasars only if the
bright end slope of the luminosity function were much steeper than 
that measured at low redshift; but such a steeper slope is not consistent 
with our sample. 
This calculation underlines the importance of searching for faint quasars
and studying the shape of the high-redshift quasar luminosity function 
at the low luminosity end. 
While it is unlikely that quasars provide the majority of photons that
ionized the universe at $z\gtrsim 6$, it still remains an open question.

\section{Summary}

In this paper, we present three new quasars at $z>5.8$, selected
from $\sim 1550$ deg$^2$ of multicolor imaging data from
the Sloan Digital Sky Survey.
The quasar candidates are selected as $i$-dropout objects, with
red $i^*-z^*$ colors. They are separated from red cool
dwarfs using follow-up $J$-band photometry.
The three new quasars, with redshifts of 5.82, 5.99 and
6.28, are the highest-redshift objects with spectral confirmation 
known to date. 

The spectra of the new quasars show very strong Ly$\alpha$ absorption
blueward of the Ly$\alpha$ emission, with more than 90\% of the flux
absorbed.
The amount of absorption increases rapidly toward $z_{abs} \sim 6$.
For the $z=6.28$ quasar SDSS 1030+0524, the flux is consistent with
zero in a region of 300\AA\ immediately blueward of Ly$\alpha$ emission.
It suggests a tentative detection of a complete Gunn-Peterson trough,
indicating that at $z\sim 6$ the universe is close to the reionization 
epoch.

The discovery spectra show tentative detection of the NV emission line,
suggesting the existence of large amount of heavy metals in the
gas around the quasars. A $J$-band spectrum of SDSS 1030+0524 shows
strong CIV emission.
The limits on the NV/CIV and NV/HeII line ratios
are consistent with the gas around the quasar having super-solar metallicity.

The three new quasars described in this paper, plus the $z=5.8$
quasar SDSS 1044--0125 (Fan et al.~2000c), form a complete color-selected,
flux-limited sample at $z\sim 6$.
We calculate the color selection completeness of this sample,
and derive the spatial density of luminous quasars at $z\sim 6$,
$\rho (M_{1450} < -26.8) = 1.1 \times 10^{-9}$ Mpc$^{-3}$, for
$\Omega =1$ and $H_0 = 50$ km s$^{-1}$ Mpc$^{-1}$.
This density is about a factor of two lower than that at $z\sim 5$,
and is consistent with an extrapolation of the observed redshift evolution 
of quasars at $3 < z < 5$.

Using the luminosity distribution of the sample, we show that the
luminosity function is shallower than 
$\Psi(L) \propto L^{-4.6}$ at the bright end. 
A larger and deeper sample is needed to better constrain the shape
of the quasar luminosity function.
We confirm that the quasar population is
unlikely to provide enough photons to ionize the universe at $z\sim6$,
unless the luminosity function is much steeper than
$\beta_{2} = -1.58$  at the faint end.

The black hole masses of these quasars are probably
several times $10^9 \rm M_{\odot}$.
The quasars are likely to reside in very massive systems,
with the minimum mass of host dark halos $\sim 10^{13} \rm M_{\odot}$.
These massive dark halos represent rare peaks in the density field at 
high redshift, and are in the steep tail of the mass function, with a
slope that is appreciably steeper than that of the quasar luminosity function.

Follow-up observations of the luminous high-redshift quasars described in 
this paper will provide excellent probes of galaxy formation and 
IGM evolution at high redshift.
In Paper II, we present high-resolution, high-S/N optical
spectroscopy of these quasars. 
These observations are compared  with detailed cosmological
simulations to constrain models of reionization in Fan et al. (2001c).  
X-ray observations (e.g., Brandt et al.~2001) will reveal
the status of the environment very close to the quasar central engine;
follow-up IR spectroscopy can provide more diagnostics on the
metallicity of the quasars and reveal the possible existence of
strong associated absorption (e.g., \cite{BAL58});
and detections of the sources at sub-millimeter wavelengths will
shed light on the possible connections between quasar activity
and starbursts (e.g., Carilli et al.~2001).

The total area of the SDSS survey is 10,000 deg$^2$.
Assuming the quasar luminosity function of Fan et al.~(2001b) and 
normalizing it to have the spatial density at $z\sim 6$ found
in this paper, we  expect to detect {\em one} $z\sim 6.6$ quasar with 
$z^* \sim 20$ ($M_{1450} \sim -28$) in the entire survey.
A quasar at this redshift has $i^* - z^* $ and $z^* - J$ colors
similar to those of early T dwarfs.
This is the highest redshift of objects we expect to find using solely
SDSS imaging data.

The Sloan Digital Sky Survey (SDSS) is a joint project 
of The University of Chicago, Fermilab, the Institute for
Advanced Study, the Japan Participation Group, 
The Johns Hopkins University, the Max-Planck-Institute for
Astronomy (MPIA), the Max-Planck-Institute for Astrophysics (MPA), 
New Mexico State University,
Princeton University, the United States Naval Observatory, 
and the University of Washington. Apache Point
Observatory, site of the SDSS telescopes, 
is operated by the Astrophysical Research Consortium (ARC). 
Funding for the project has been provided by the 
Alfred P. Sloan Foundation, the SDSS member institutions,
the National Aeronautics and Space Administration, 
the National Science Foundation, the U.S. Department of
Energy, the Japanese Monbukagakusho, 
and the Max Planck Society. The SDSS Web site is
http://www.sdss.org.
This publication makes use of data products from the Two Micron All Sky Survey,
which is a
joint project of the University of Massachusetts and the Infrared Processing and
 Analysis
Center/California Institute of Technology, funded by NASA and NSF.
We thank the staff at APO, Keck, Calar Alto, and UKIRT, 
for their expert help,
Gemini staff scientist Marianne Takamiya for
carrying out the Germini/NIRSPEC service observation.
We thank Ed Turner, Peng Oh, Julian Krolik, Kristian Finlator, 
John Bahcall and an anonymous referee for helpful discussions and comments.
We acknowledge support from NSF grant PHY00-70928 and 
a  Frank and Peggy Taplin Fellowship (XF), NSF grant
AST-0071091 (MAS), and NSF grant AST-9900703 (DPS).

\newpage

\begin{deluxetable}{rrrrcc}
\tablenum{1}
\tablecolumns{6}
\tablecaption{Summary of Photometric Runs}
\tablehead
{
run & strip & $\lambda_{min}$ & $\lambda_{max}$ & $\langle$ FWHM($i$) $\rangle$ ($''$) & $\langle$ FWHM($z$)  $\rangle$ ($''$)
}
\startdata
 745 & 10 N &  51.1 &  65.0 & 1.23 & 1.25 \\ 
 752 & 10 S & $-17.6$ &  51.0 & 1.43 & 1.50 \\ 
 756 & 10 N & $-63.1$ &  51.0 & 1.34 & 1.38 \\ 
1140 &  9 N & $-30.5$ &  17.1 & 1.44 & 1.47 \\ 
1231 &  9 S & $-14.2$ &  14.7 & 1.51 & 1.53 \\ 
1239 & 10 S & $-63.1$ & $-39.7$ & 1.41 & 1.46 \\ 
1241 &  9 S & $-45.0$ & $-38.9$ & 1.49 & 1.57 \\ 
1241 &  9 S & $-24.8$ & $-19.0$ & 1.74 & 1.76 \\ 
1331 & 36 S & $-56.4$ & $-36.0$ & 1.18 & 1.18 \\ 
1331 & 36 S & $-30.2$ & $-18.0$ & 1.38 & 1.36 \\ 
1332 & 36 S & $-18.1$ & $-10.2$ & 1.52 & 1.46 \\ 
1336 & 42 N &  23.4 &  36.1 & 1.51 & 1.50 \\ 
1339 & 42 S &  23.3 &  36.0 & 1.41 & 1.38 \\ 
1345 & 36 N & $-52.6$ & $ -8.0$ & 1.42 & 1.42 \\ 
1345 & 36 N &   1.3 &  73.9 & 1.23 & 1.26 \\ 
1350 & 37 S & $-53.7$ &  14.5 & 1.54 & 1.54 \\ 
1356 & 43 N &  22.8 &  36.5 & 1.64 & 1.62 \\ 
1359 & 43 S &  27.7 &  37.2 & 1.61 & 1.57 \\ 
1402 & 37 N & $-51.0$ & $-34.1$ & 1.24 & 1.22 \\ 
1412 & 37 N & $-23.6$ &  33.2 & 1.28 & 1.28 \\ 
1449 &  1 N &   2.9 &  13.4 & 1.46 & 1.51 \\ 
1450 & 37 N & $-34.2$ & $-23.3$ & 1.53 & 1.54 \\ 
1462 & 11 S & $-56.1$ &  40.6 & 1.27 & 1.35 \\ 
1468 & 32 N &  10.4 &  29.1 & 1.37 & 1.35 \\ 
1469 & 32 S &   7.5 &  29.4 & 1.49 & 1.48 \\ 
1478 & 12 S &  25.6 &  55.8 & 1.43 & 1.49 \\ 
2074 & 35 S & $-34.6$ & $-25.2$ & 1.51 & 1.56 \\ 
2076 & 35 N & $-56.2$ & $-33.9$ & 1.33 & 1.35 \\ 
2078 & 36 S & $ -9.4$ &   8.5 & 1.47 & 1.49 \\ 
2125 & 12 S & $-64.5$ & $-17.1$ & 1.53 & 1.58 \\ 
2126 & 12 N & $-64.2$ &   0.2 & 1.30 & 1.36 \\ 
2131 & 34 S & $-60.3$ & $-42.6$ & 1.41 & 1.50 \\ 
2134 & 36 S &   8.4 &  23.5 & 1.48 & 1.41 \\ 
2137 & 34 N & $-59.3$ & $-19.6$ & 1.37 & 1.43 \\ 
2138 & 34 N & $-22.7$ & $ -9.5$ & 1.51 & 1.55 \\ 
2140 & 34 S & $-45.4$ & $-35.3$ & 1.69 & 1.76 \\ 
2141 & 10 S & $-38.0$ & $-17.2$ & 1.42 & 1.51 \\ 
2190 & 12 N &   2.3 &  28.2 & 1.47 & 1.54 \\ 
2190 & 12 N &  32.9 &  51.7 & 1.48 & 1.53 
\enddata
\end{deluxetable}

\newpage
\begin{deluxetable}{rrr}
\tablenum{2}
\tablecolumns{3}
\tablecaption{Summary of Follow-up Results}
\tablehead
{
  & number of objects & percentage 
}
\startdata
$z>5.8$ quasars & 4 & 3.3\% \\
T dwarfs & 11 & 9.1\% \\
M/L dwarfs by spectroscopy & 10 & 8.3\% \\
M/L dwarfs by photometry & 60 & 49.6 \% \\
BAL quasar & 1 & 0.8\% \\
false detection & 35 & 28.9\% \\ \hline
TOTAL & 121 & 100\%
\enddata
\end{deluxetable}

\begin{deluxetable}{cccccc}
\tablenum{3}
\tablecolumns{6}
\tablecaption{Photometric Properties of Three New $z>5.8$ Quasars}
\tablehead
{
object & redshift & $i^*$ & $z^*$ & $J$ & SDSS run
}
\startdata
J083643.85+005453.3 & 5.82 $\pm$ 0.02  & 21.04 $\pm$ 0.08  & 18.74 $\pm$ 0.05  & 17.89 $\pm$ 0.05  & 1239 \\ 
J103027.10+052455.0 & 6.28 $\pm$ 0.03  & 23.23 $\pm$ 0.43  & 20.05 $\pm$ 0.10  & 18.87 $\pm$ 0.10  & 2125 \\ 
J130608.26+035626.3 & 5.99 $\pm$ 0.03  & 22.58 $\pm$ 0.26  & 19.47 $\pm$ 0.05  & 18.77 $\pm$ 0.10  & 2190 
\enddata

\tablenotetext{}{The SDSS photometry ($i^*,z^*$) is
reported in terms of {\em asinh magnitudes} on the AB system.
The asinh magnitude system is defined by Lupton, Gunn \& Szalay (1999);
it becomes a linear scale in flux when the absolute value of the
signal-to-noise ratio is less than about 5. In this
system, zero flux corresponds to 24.4 and 22.8,
in $i^*$, and $z^*$, respectively; larger magnitudes refer to negative flux values.
The $J$ magnitude is on the Vega-based system.}
\end{deluxetable}

\newpage
\begin{deluxetable}{ccccccc}
\tablenum{4}
\tablecolumns{7}
\tablecaption{Average Ly$\alpha$ Absorption of $z>5.8$ Quasars}
\tablehead
{
object & redshift & $D_{A}$ & $D_B$ & $z_{abs}$ & Transmitted flux
ratio\\
&&&&& at $z_{abs}$ 
}
\startdata
J104433.04--012502.2 & 5.80 & 0.91 $\pm$ 0.02 & 0.95 $\pm$ 0.02 &  5.5 & 0.088 $\pm$ 0.004 \\
J083643.85+005453.3 & 5.82 & 0.90 $\pm$ 0.02 & 0.91 $\pm$ 0.02 & 5.5 & 0.099 $\pm$ 0.014 \\
J130608.26+035626.3 & 5.99 & 0.92 $\pm$ 0.02 & 0.95 $\pm$ 0.02 & 5.7 & 0.069 $\pm$ 0.014 \\
J103027.10+052455.0 & 6.28 & 0.93 $\pm$ 0.02 & 0.99 $\pm$ 0.01 & 6.0 & 0.003 $\pm$ 0.020
\enddata
\end{deluxetable}

\begin{deluxetable}{ccccccc}
\tablenum{5}
\tablecolumns{7}
\tablecaption{Continuum Properties of $z>5.8$ Quasars in the Complete Sample}
\tablehead
{
object & redshift & $m_{1280}$ & $m_{1450}$ & $M_{1450}$ ($\Omega$-model) & $M_{1450}$ ($\Lambda$-model) & $E(B-V)$
}
\startdata
J083643.85+005453.3 & 5.82 & 18.88 & 18.81 & --27.62 & --27.88 & 0.050 \\ 
J103027.10+052455.0 & 6.28 & 19.73 & 19.66 & --26.89 & --27.15 & 0.023 \\ 
J104433.04--012502.2 & 5.80 & 19.28 & 19.21 & --27.15 & --27.50 & 0.054 \\ 
J130608.26+035626.3 & 5.99 & 19.61 & 19.55 & --26.93 & --27.19 & 0.028 
\enddata
\end{deluxetable} 

\newpage

\begin{figure}
\vspace{-2.2cm}

\epsfysize=600pt \epsfbox{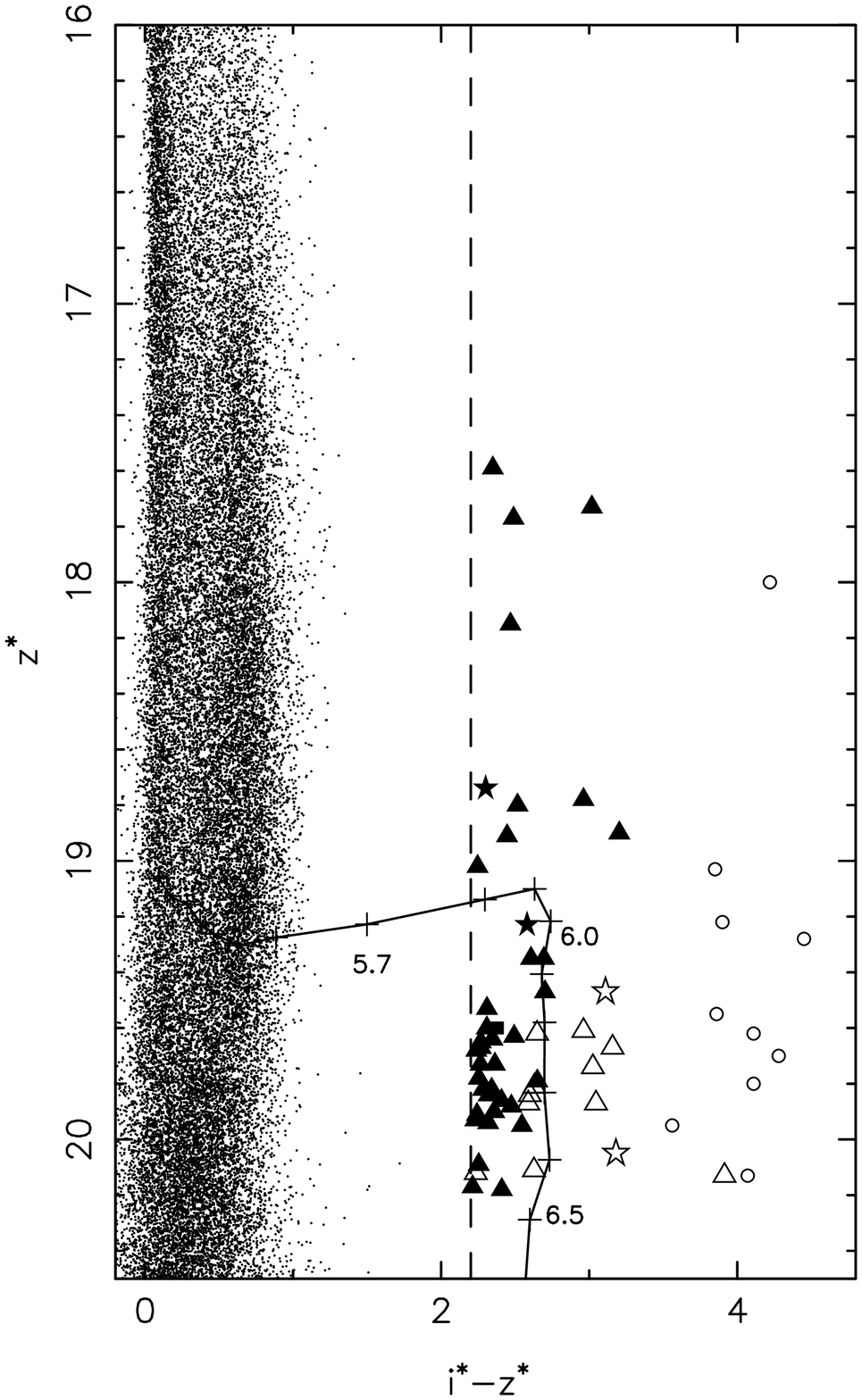}

Figure 1. $i^*-z^*$ vs.~$z^*$ color-magnitude diagram for the
$i$-dropout sample. Colors and magnitudes are plotted as
asinh magnitudes measured by SDSS imaging.
The symbols represent the classification
in Table 2: circles: T dwarfs; triangles: L dwarfs; 
stars: $z>5.8$ quasars; square: BAL quasar. Filled symbols are objects
with S/N in the $i$ band higher than 4; open symbols are objects not
detected in the $i$ band at the 4-$\sigma$ level.
The median track of simulated $i^*-z^*$ color and $z^*$ magnitude for
quasars with $M_{1450} = -27$ is also shown as a function of redshift,
with plus signs every 0.1 in redshift.
For comparison, the data for a random sample of 50,000 high-latitude
stars are also shown as dots. The dashed line shows the cut
$i^* - z^* > 2.2$ that we use to select high-redshift quasar
candidates. 

\end{figure}

\begin{figure}
\vspace{-2cm}

\epsfysize=600pt \epsfbox{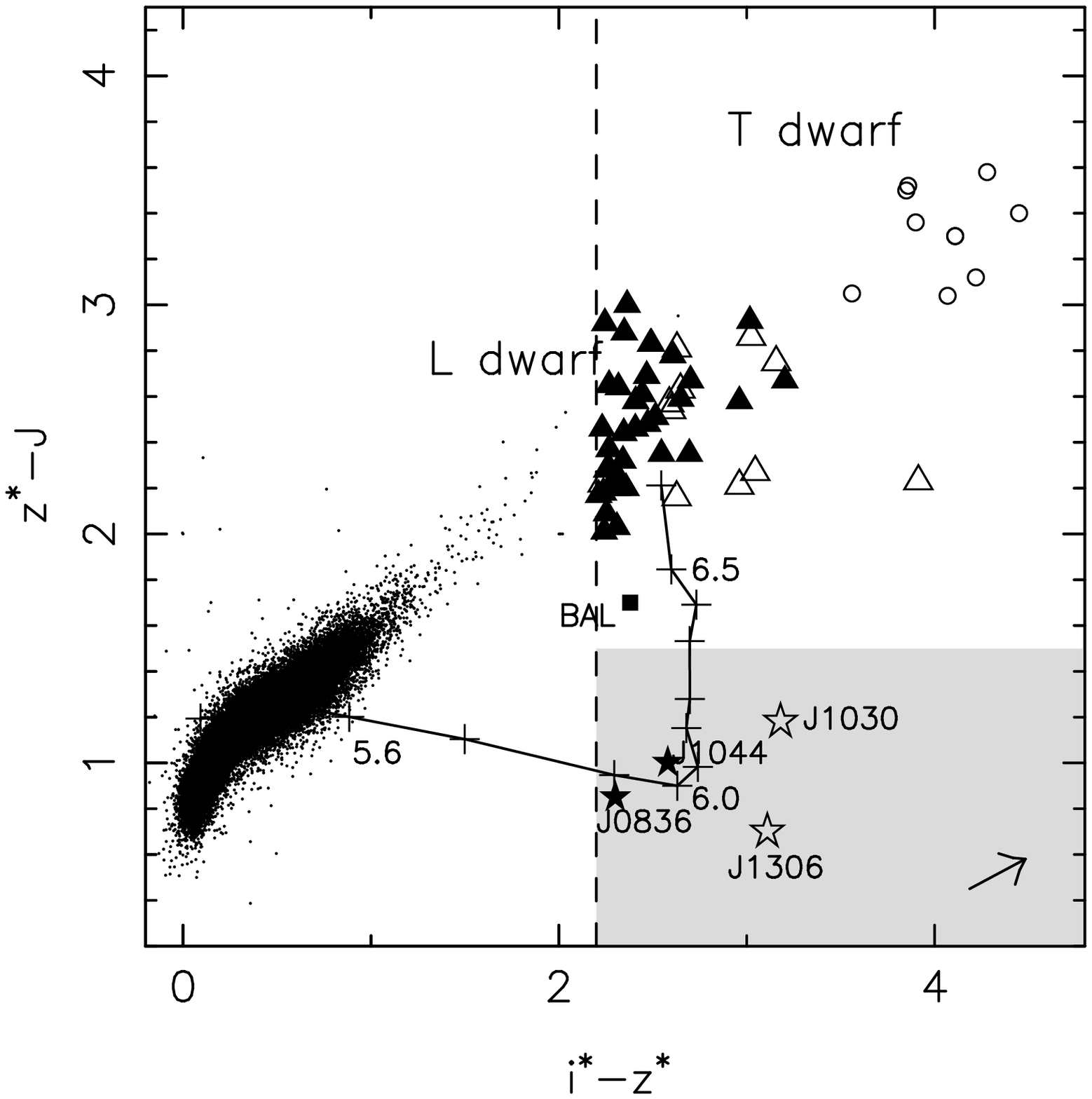}

\vspace{1cm}
Figure 2. $i^*-z^*$ vs. $z^*-J$ color-color diagram for the 
$i$-dropout sample. The symbols are the same as in Figure 1.
The median track of simulated quasar colors is shown as a function
of redshift.
The survey selection criteria are illustrated by the shaded area.
For comparison, colors of SDSS-2MASS stars in a 50  deg$^2$
area at high latitude are also shown.
The arrow at the lower-right corner indicates the reddening
vector for quasars at $z=6$ with $E(B-V) = 0.1$.

\end{figure}

\begin{figure}
\vspace{-3cm}

\epsfysize=600pt \epsfbox{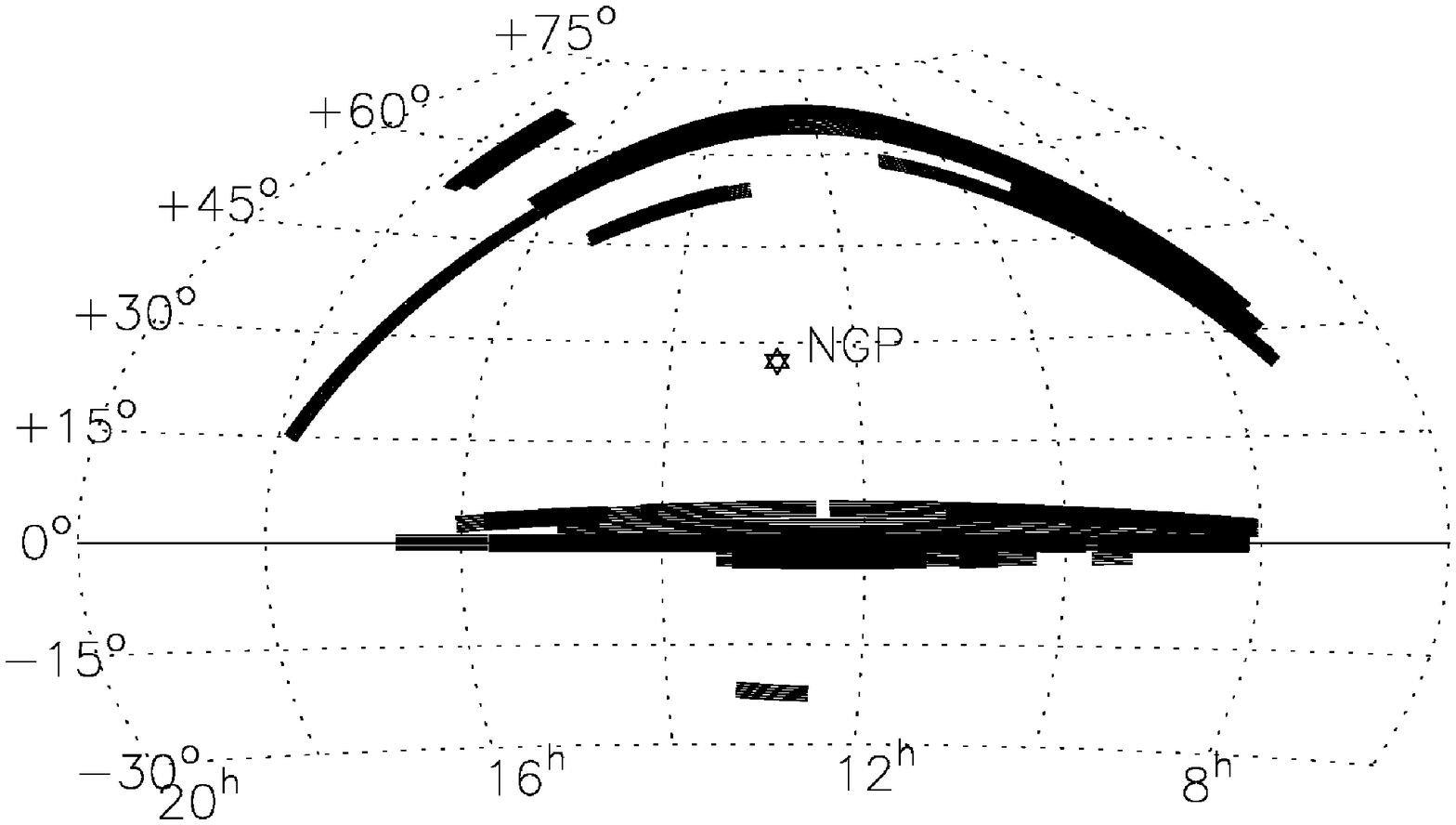}

\vspace{1cm}
Figure 3. Projection of the survey area in J2000 equatorial coordinates.
Note that for some SDSS stripes, only one of the two scans is covered
by the current survey. The stripes are each $2.5^\circ$ wide. 

\end{figure}

\begin{figure}
\vspace{-3cm}

\epsfysize=600pt \epsfbox{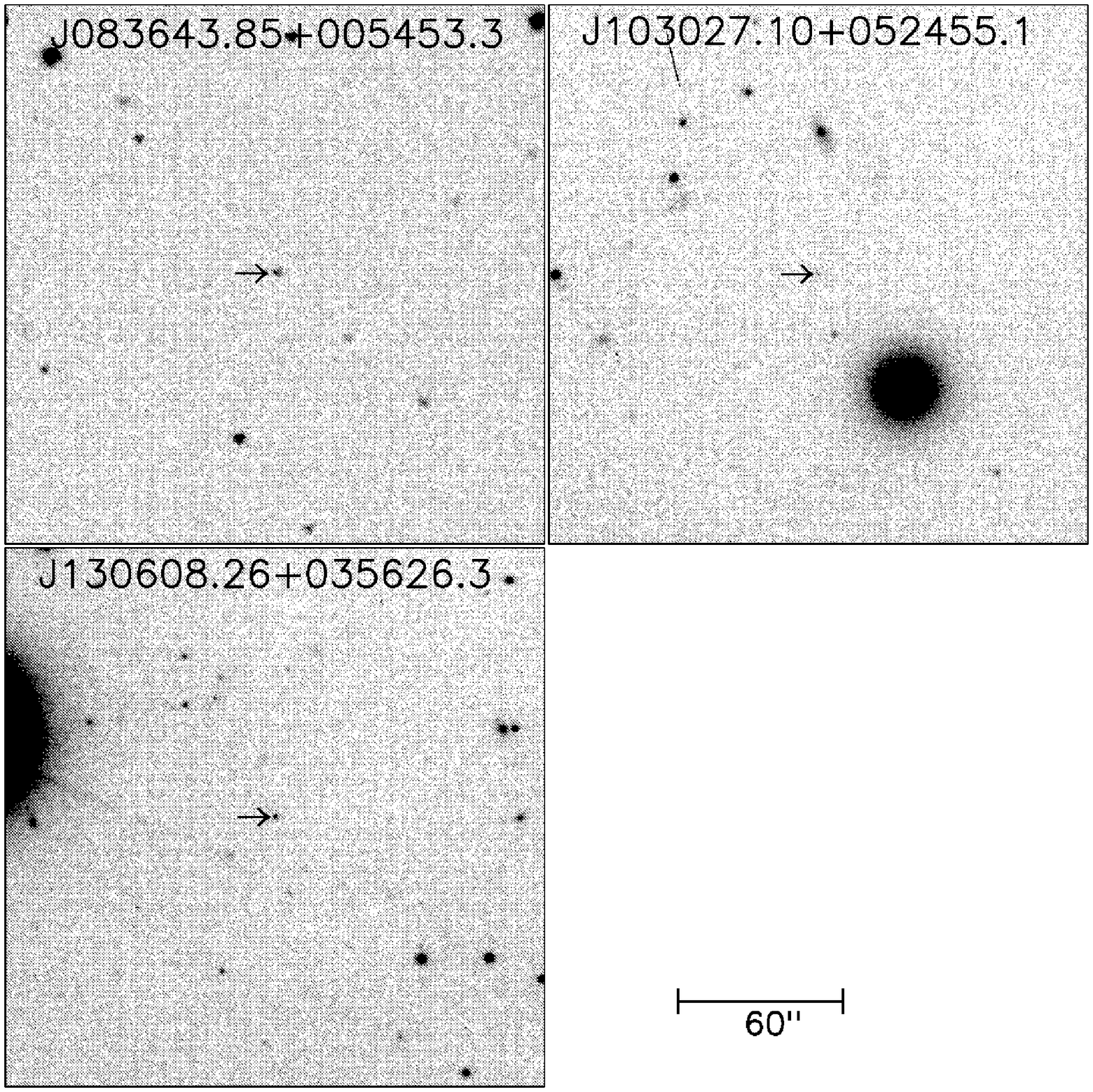}

\vspace{1cm}
Figure 4. $z$ finding charts of the three new $z>5.8$ quasars.
For SDSS 0836+0054, the SDSS $z$ image is shown.
For SDSS 1030+0524 and SDSS 1306+0356, the ARC 3.5m $z$ images taken with
SPICAM are shown (180 seconds exposure).
The size of the finding chart is 160''. North is up and East is left.

\end{figure}

\begin{figure}
\vspace{-1cm}
 
\epsfysize=600pt \epsfbox{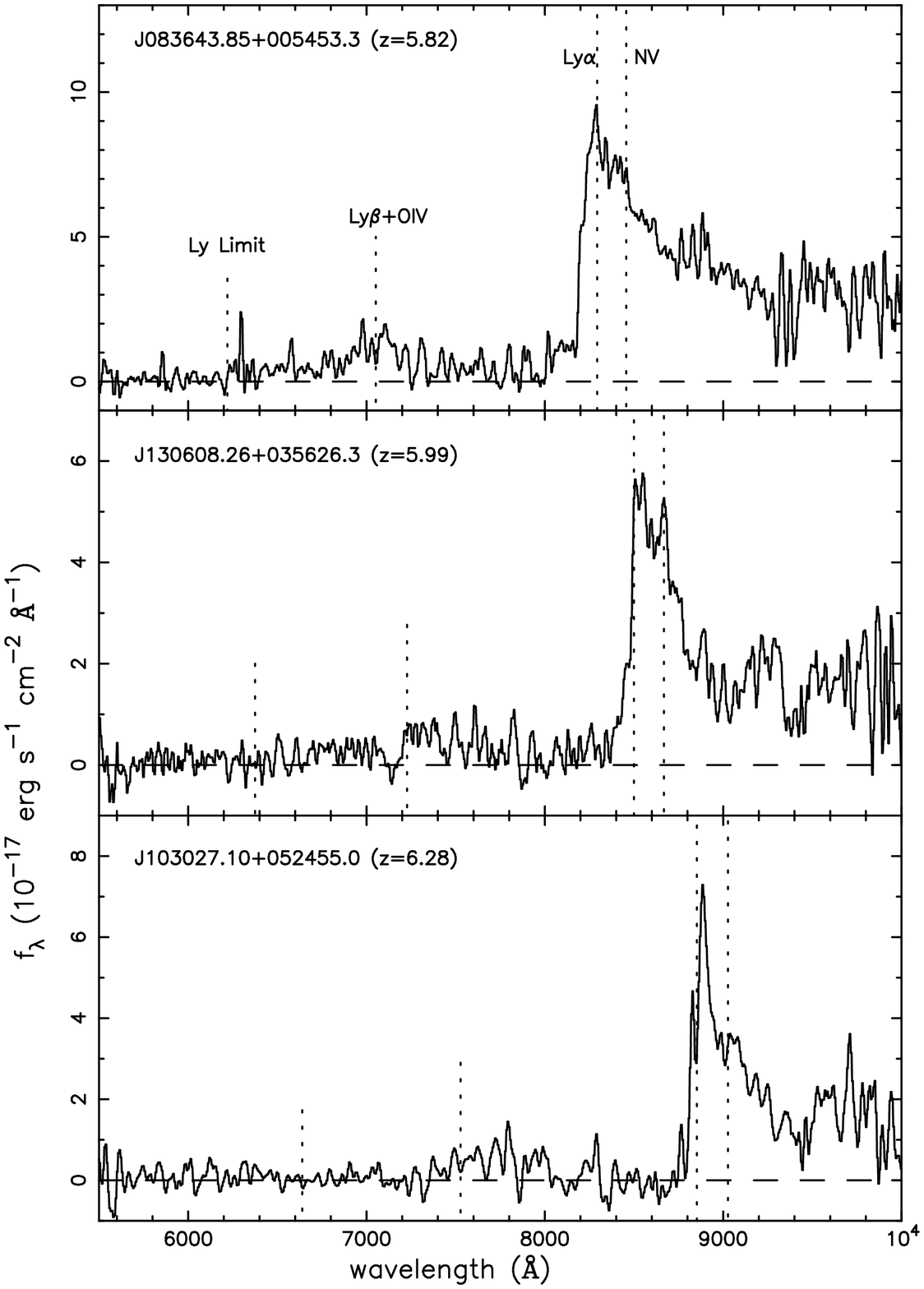}

\vspace{0.2cm}
Figure 5. The discovery spectra of the three new $z>5.8$ quasars,
taken with the ARC 3.5m telescope and DIS spectrograph.
The exposure time is 3600 seconds for each quasar and the spectral
resolution is about 20\AA.  The spectra have been normalized to the
photometry in the $z$ filter. 

\end{figure}
\begin{figure}
\vspace{-2cm}

\epsfysize=600pt \epsfbox{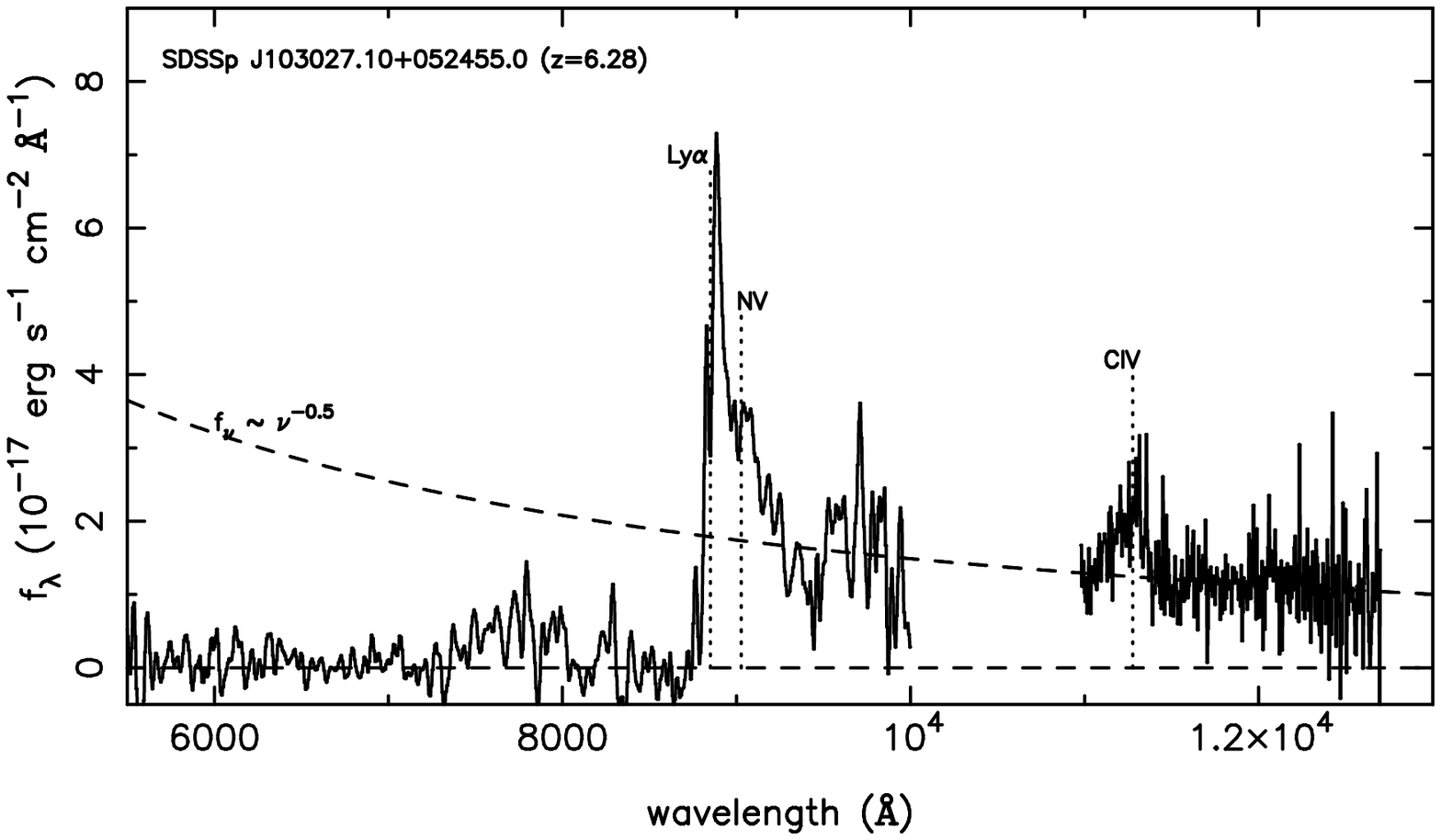}

\vspace{1cm}
Figure 6. Combined optical + near IR  spectrum of SDSS 1030+0524.
The optical spectrum is  a 3600 second exposure taken with ARC 3.5m telescope (same as in Figure 5).
The near IR spectrum is a 3000 second exposure taken with Keck/NIRSPEC.
The resolution of the near IR spectrum is $R\sim 1500$.
\end{figure}
\begin{figure}
\vspace{-2cm}

\epsfysize=600pt \epsfbox{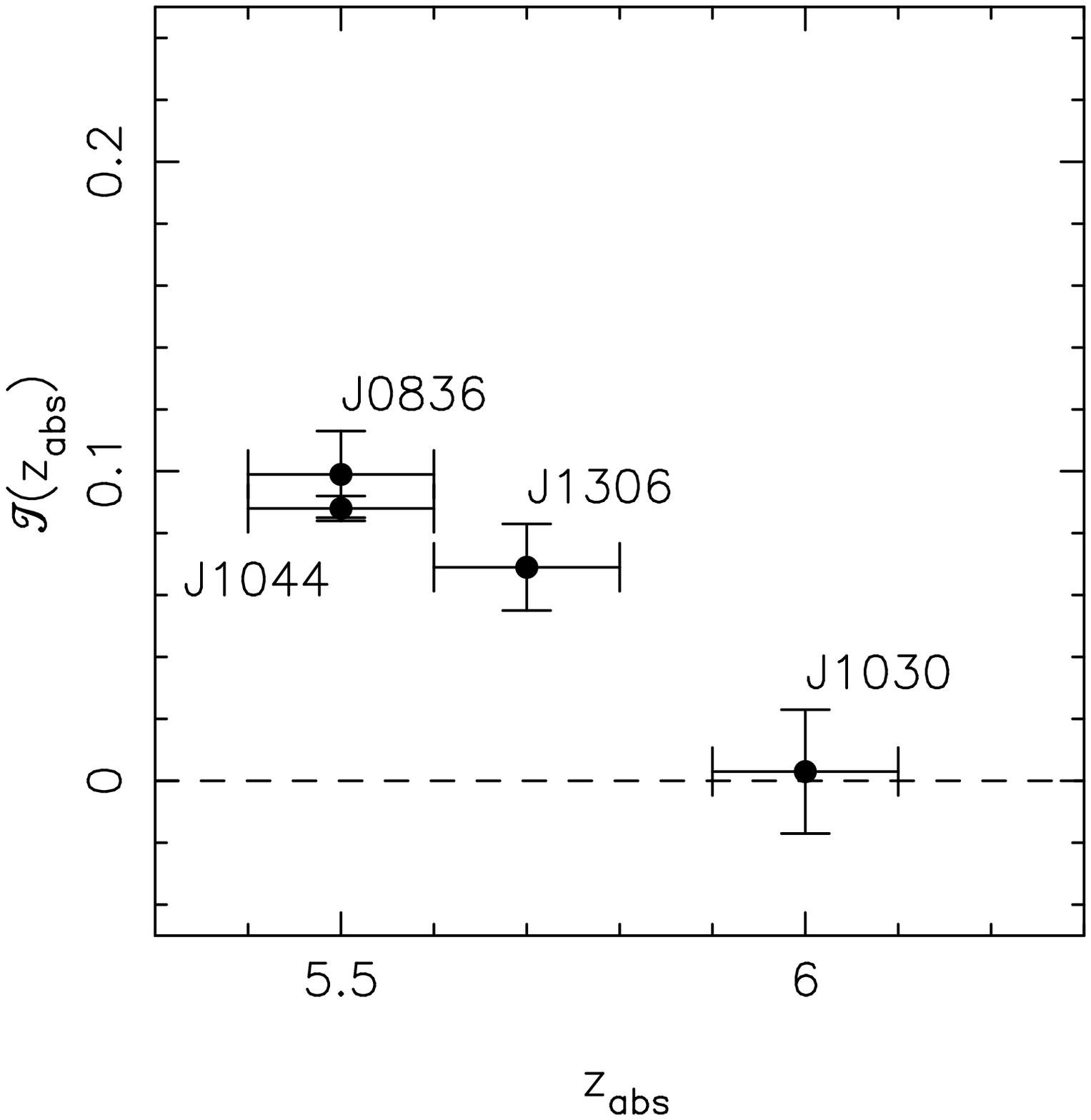}

\vspace{1cm}
Figure 7. Evolution of the transmitted flux ratio in the Ly$\alpha$
forest region, averaged over the wavelength range
$(1+z_{abs} - 0.1)\times 1216$\AA\ $\lesssim \lambda \lesssim (1+z_{abs}
+ 0.1)\times 1216$\AA. The vertical error bars represent the photon noise. Note that 
for SDSS 1030+0524, the flux is consistent with zero in the Ly$\alpha$ region
immediately blueward of the Ly$\alpha$ emission.  

\end{figure}

\begin{figure}
\vspace{-1cm}
 
\epsfysize=600pt \epsfbox{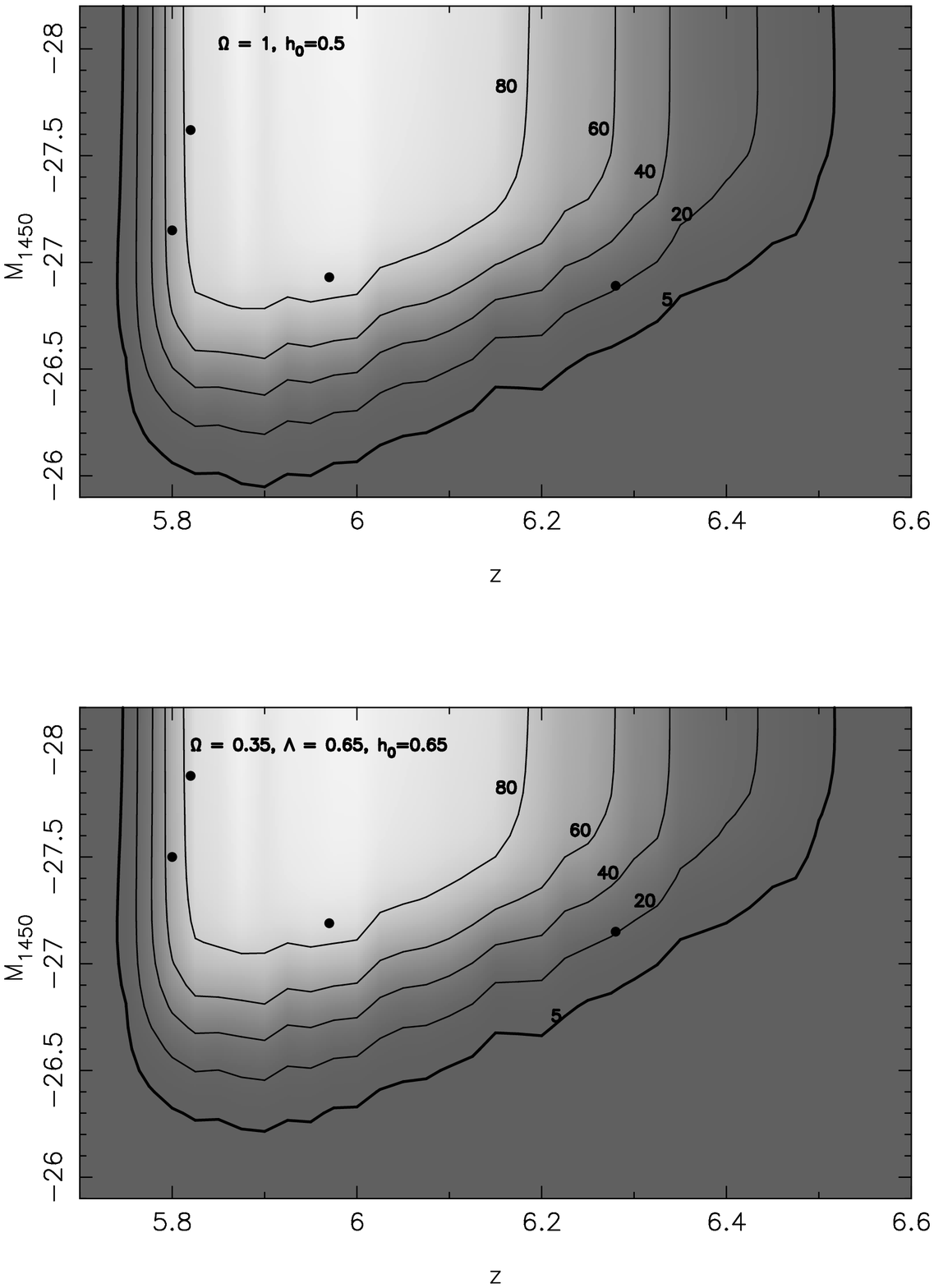}

\vspace{1cm}
Figure 8. The selection probability of $z>5.8$ quasars
as a function of redshift and
luminosity, for the $\Omega=1$ and the $\Lambda$ models.
The probability contours of 5\%, 20\%, 40\%, 60\% and 80\% are shown.
The large dots represent the locations of the 4 quasars in the sample.

\end{figure}

\begin{figure}
\vspace{-3cm}
 
\epsfysize=600pt \epsfbox{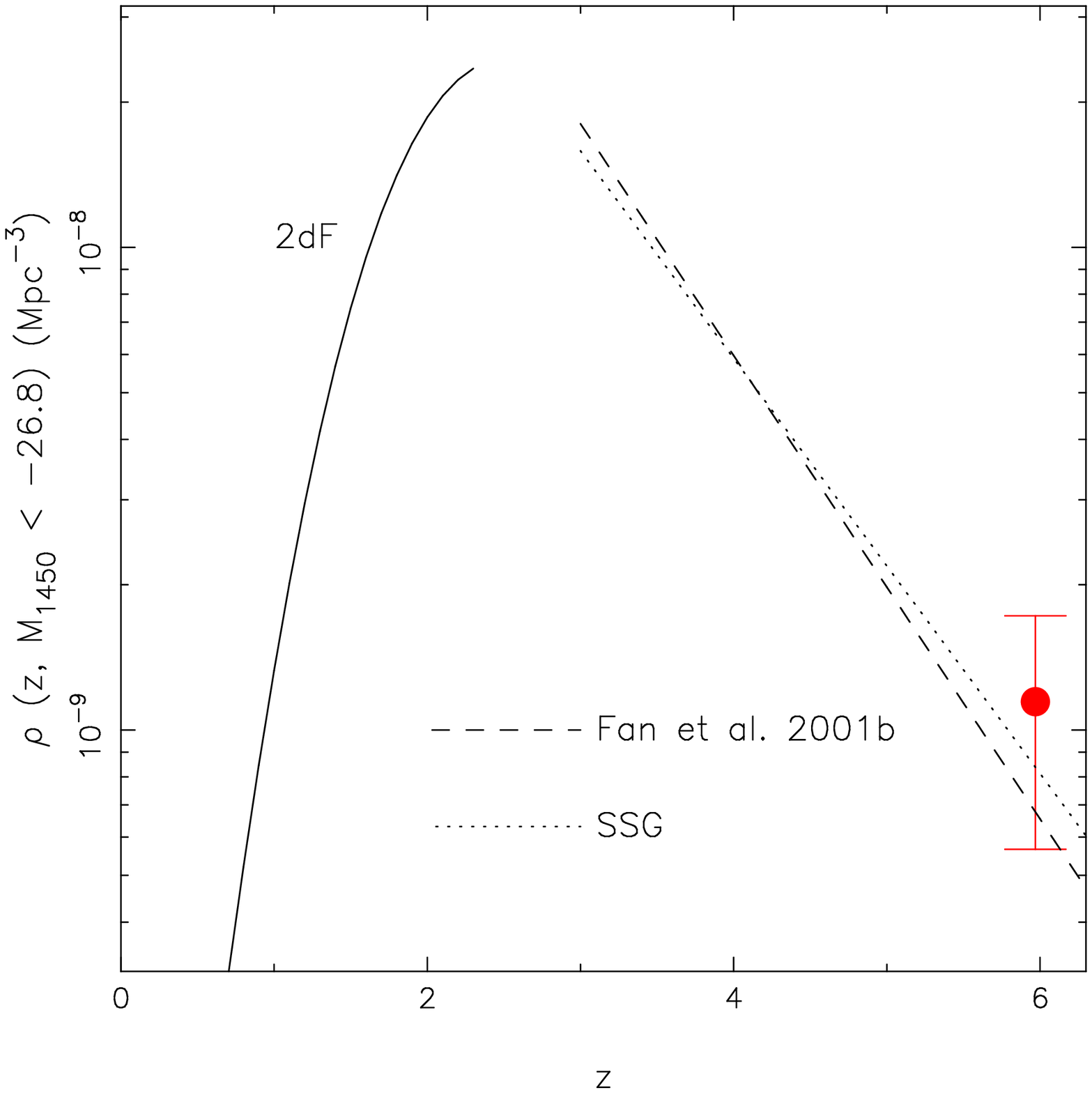}

\vspace{1cm} 

Figure 9. The evolution of quasar comoving spatial density at 
$M_{1450} < -26.8$ in the $\Omega=1$ model.
The large dot represents the result from this survey.
The dashed and dotted lines are the best-fit models
from Fan et al.~(2001b) and Schmidt et al.~(1995),
respectively.
The solid line is the best-fit model from the 2dF survey
(Boyle et al.~2000) at $z<2.5$.
\end{figure}

\begin{figure}
\vspace{-3cm}

\epsfysize=600pt \epsfbox{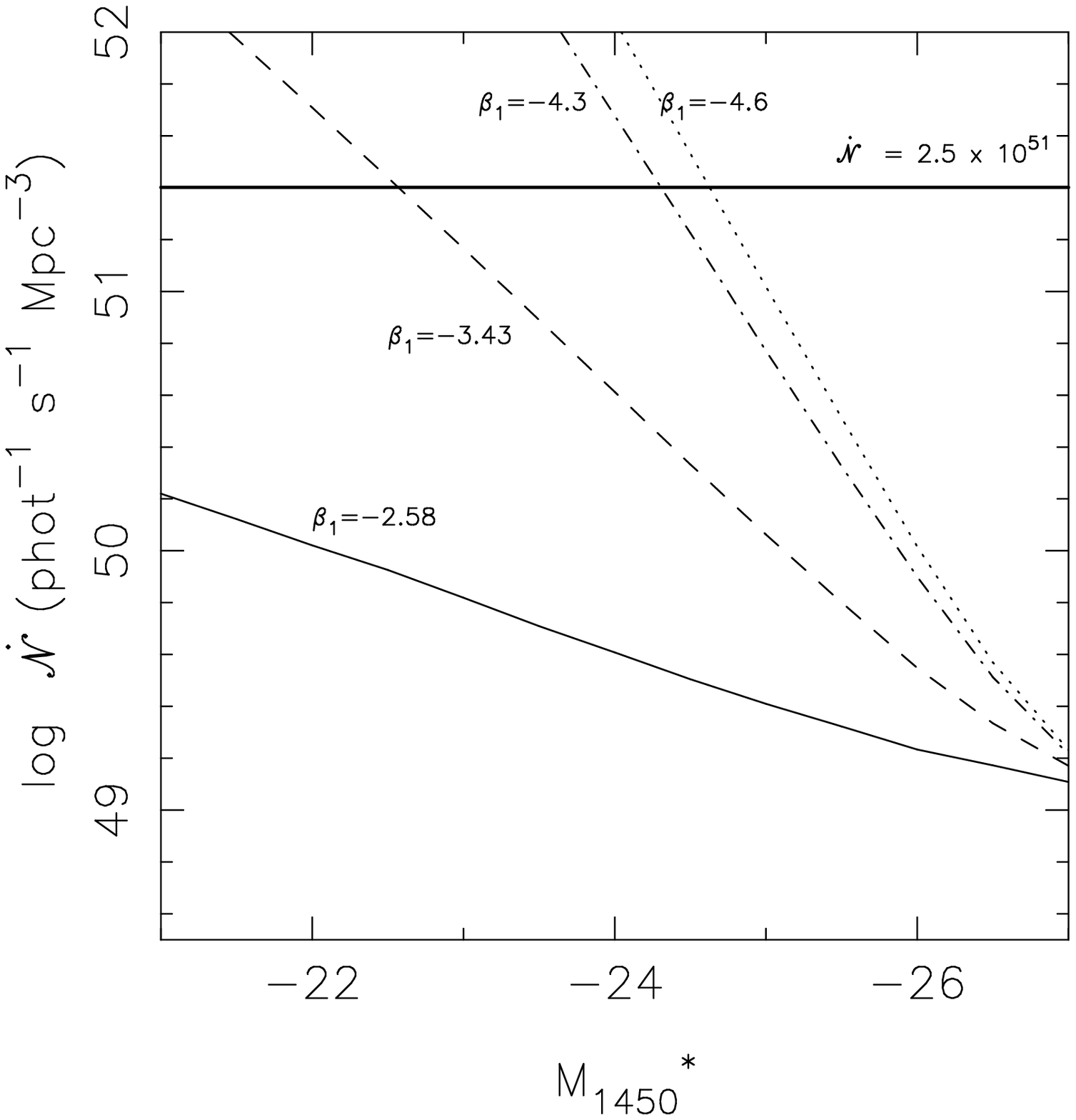}

\vspace{1cm}

Figure 10. Comoving emission density rate of hydrogen Lyman continuum
photons from quasars integrated over the full luminosity function,
compared with the minimum rate (heavy solid line) that is needed to
fully ionize a universe at $z=6$, in the $\Omega=1$ model, with a
clumping factor of 30, and the baryon fraction $\Omega_b h^2 = 0.02$.
The recombination time is assumed to be faster than the Hubble time.
The quasar emission rate is calculated for different bright end slopes
of the quasar luminosity function ($\beta_1$) and turnover luminosity
$M_{1450}^\ast$, assuming the same faint end slope $\beta_2 = -1.58$
and the total density of luminous quasars matched to that found in
this paper, $\rho (z=6, M_{1450} < -26.8) =1.14 \times 10^{-9}$
Mpc$^{-3}$. 
\end{figure}

\end{document}